\documentclass [12pt,a4paper]{article}

\usepackage{graphicx}
\usepackage{epsfig}
\usepackage{amsfonts}
\renewcommand{\vec}[1]{\mbox{\boldmath$#1$}}

\newcommand{\eps}{\varepsilon}

\begin{document}

\title{\textbf{Finite Volume Analysis of Nonlinear Thermo-mechanical
Dynamics of \\ Shape Memory Alloys}}

\author{L. X.  Wang$^1$ \quad and \quad
Roderick V.N. Melnik$^2$\footnote{Corresponding author: tel: +1-519-884-1970, fax:
 +1-519-884-9738, email: rmelnik@wlu.ca} \\
 $^1$MCI, Faculty of Science and Engineering, \\
 University of Southern Denmark,\\
 Sonderborg, DK-6400, Denmark \\
$^2$Mathematical Modelling and Computational Sciences, \\
Wilfrid Laurier University, 75 University Ave West, \\
Waterloo, ON, Canada N2L 3C5
}
\date{}

\maketitle


\begin{abstract}
In this paper, the finite volume method is developed to analyze coupled dynamic problems of
nonlinear thermoelasticity. The major focus is given to the description of martensitic
   phase transformations
essential in the modelling of shape memory alloys. Computational experiments are carried out
to study the thermo-mechanical wave interactions in a shape memory alloy rod, and a patch.
Both mechanically and thermally induced phase transformations, as well as hysteresis effects,
in a one-dimensional structure are successfully simulated with the developed methodology. In
the two-dimensional case, the main focus is given to square-to-rectangular transformations
and examples of martensitic combinations under different mechanical loadings are provided.

\vspace*{0.3cm}
\end{abstract}


\section{Introduction}

The existing and potential applications of Shape Memory Alloys (SMA) lead to
an increasing interest to the analysis of these materials by means of
 both experimental and
theoretical approaches \cite{Birman1997}. These materials have unique properties thanks to
their unique ability to undergo reversible phase transformations when subjected to
appropriate thermal and/or mechanical loadings. Mathematical modelling tools play an
important role in studying such transformations and computational experiments, based on
mathematical models, can be carried out to predict the response of the material under various
loadings, different types of phase transformations, and reorientations.  The development of
such tools is far from straightforward even in the one-dimensional case where
 the analysis of the dynamics is
quite involved due to a strongly nonlinear pattern of interactions between mechanical and
thermal fields (e.g., \cite{Birman1997,Matus2004} and references therein).  For a number of
practical applications a better understanding of the dynamics of SMA structures with
dimensions higher than one becomes critical. This makes the investigation more demanding,
both
 theoretically and numerically.

Most results reported so far for the one-dimensional case have been obtained with the Finite
Element Method (FEM) \cite{Bubner1996,Bubner2000,Niezgodka1991}. In addition to the
challenges pertinent to coupling effects, we have to deal also with strong nonlinearities of
the problem at hand. One of the approaches is to employ a FEM using cubic spline basis
functions, in which case the nonlinear terms can be smoothed out by one of the available
averaging algorithms. As an explicit time integration is typically employed in such
situations, this results in a very small time step discretization.
 Seeking for a more efficient numerical
approach, Melnik et al. \cite{Melnik2000,Melnik2002} used a differential-algebraic
methodology to study the dynamics of martensitic transformations in a SMA rod. An extension
of that approach has been recently developed in \cite{Matus2003,Matus2004,Melnik2003} where
the authors constructed a fully conservative, second-order finite-difference scheme that allowed
them to carry out computations on a minimal stencil. However, a direct generalization of the
scheme to a higher dimensional case appeared to be difficult.

In this paper, we approach the same problem from the Finite Volume Method (FVM) point of
view. The method is based on the integral form of the governing equations, leading to
inherently
 conservative properties of FVM numerical schemes. The methodology is well
 suited for treating complicated, coupled multiphysics nonlinear
problems \cite{Berezovski2003,Demirdzic1994,Demirdzic1997}. It can be relatively easily
generalized to higher dimensional cases. In addition to its wide-spread popularity in CFD,
the method has been applied previously to linear elastic and thermoelastic problems
\cite{Berezovski2001,Demirdzic1994,Demirdzic1997,Jasak2000}. There are several recent results
on the application of FVM to nonlinear thermo-mechanical problems and nonlinear elastic
problems \cite{Berezovski2003,Tuzel2004}. In this paper, we develop a FVM specifically in the
context of studying martensitic transformations in SMAs and demonstrate its performance in
simulating the dynamical behavior of SMA rods and patches.

The paper is organized as follows. The mathematical models for the dynamics of martensitic
transformations in 1D and 2D SMA structures are described in Section 2. Key issues  of
numerical discretization of these models, including the FVM and its computational
implementation via the
 Differential-Algebraic Equations (DAE) approach,
 are discussed in Section 3. Mechanically and thermally induced transformations and
hysteresis effects in SMA rods are analyzed in Section 4. Section 5 is devoted to studying
nonlinear thermomechanical behavior  and square-to-rectangular transformations in a SMA
patch. Finally, conclusions are given in Section 6.


\section{Mathematical Model for SMA Dynamics}

We start our consideration from a  mathematical model for the SMA dynamics
 based on a coupled system of the three fundamental laws, conservation of mass, linear momentum,
 and energy balance,  in a way we described previously in  \cite{Melnik2003,Matus2004,Wang2004}. Using these laws, the
system that describes coupled thermo-mechanical wave interactions for the first
order martensitic phase transformations in a three dimensional SMA structure can
be written as follows \cite{Melnik2000, Melnik2003,Pawlow2000}
\begin{equation}
\label{eq2-1}
\begin{array}{l}  \displaystyle
 \rho\frac{\partial^{2}u_{i}}{\partial t^{2}} =
\nabla_{x}\cdot\vec{\sigma}+f_{i},\quad i,j=1,2,3
\\[10pt] \displaystyle
\rho\frac{\partial e}{\partial t } -
 \vec{\sigma}^{T}:\nabla\mathbf{v}+\nabla\cdot
\mathbf{q}=g  ,
\end{array}
\end{equation}
where $\rho$ is the density of the material, $\textbf{u}=\{
u_{i}\}|_{i=1,2,3}$ is the displacement vector, $\textbf{v}$ is the velocity,
$\vec{\sigma}=\{\sigma_{ij}\}$ is the stress tensor, $\textbf{q}$ is the heat
flux, $e$ is the internal energy, $\textbf{f}=(f_{1},f_{2},f_{3})^{T}$ and $g$
are distributed mechanical and thermal loadings, respectively. Let $\phi$ be the
free energy function of a thermo-mechanical system described by (\ref{eq2-1}),
then, the stress and the internal energy function are connected with $\phi$ by
the following relationships:

\begin{equation}
\label{eq2}  \displaystyle
\vec{\sigma}= \frac{\partial\phi}{\partial\vec{\eta}},
\quad{e=\phi-\theta\frac{\partial\phi}{\partial\theta}},
\end{equation}
where $\theta$ is the temperature, and $\vec{\eta}$ the Cauchy-Lagrangian
strain tensor defined as follows:
\begin{equation}
\label{eq2-3}  \displaystyle
\eta_{ij}\left(\textrm{\textbf{x}},t\right)=\left(\frac{\partial
u_{i}\left(\textrm{\textbf{x}},t\right)}{\partial x_{j}}+\frac{\partial
u_{j}\left(\textbf{x},t\right)}{\partial x_{i}}\right)/2.
\end{equation}

In what follows, we employ the Landau-Ginzburg form of the free energy function for both 1D
and 2D SMA dynamical models \cite{Bubner1996,Falk1980,Melnik2000}. In the 2D case, we focus
 our attention on the square-to-rectangular transformations that can be regarded as a 2D analog of the
realistic cubic-to-tetragonal and tetragonal-to-orthorhombic transformations
\cite{Ichitsubo2000,Jacobs2000}. It is known that  for this kind of transformations, the free
energy function $\phi$ can be constructed by taking advantage of a Landau free energy
function $F_{L}$. In particular, following \cite{Ichitsubo2000,Jacobs2000,Lookman2003} (see
also references therein), we have:
\begin{equation}
\label{eq2-4}  \displaystyle
\phi=-c_{v} \theta \ln \theta + \frac{1}{2} a_{1}
e_{1}^{2}+ \frac{1}{2}a_{3}e_{3}^{2}+F_{L},\quad
F_{L}=\frac{1}{2}a_{2}\left(\theta-\theta_{0}\right)e_{2}^{2}-\frac{1}{4}a_{4}
e_{2}^{4}+\frac{1}{6}a_{6}e_{2}^{6},
\end{equation}
where $c_{v}$ is the specific heat constant, $\theta_0$ is the reference
temperature for the martensite transition, $a_{i}$, $i=1,2,3,4,6$ are the
material-specific coefficients, and $e_{1}$, $e_{2}$, $e_{3}$ are dilatational,
deviatoric, and shear components of strain, respectively. The latter are defined
as follows:
\begin{equation}
\label{eq2-5}  \displaystyle
e_{1}=\left(\eta_{11}+\eta_{22}\right)/\sqrt{2},\quad
e_{2}=\left(\eta_{11}-\eta_{22}\right)/\sqrt{2},\quad
e_{3}=\left(\eta_{12}+\eta_{21}\right)/2.
\end{equation}
This free energy function $\phi$ is a convex function of the chosen order parameters
when the temperature is much higher than
$\theta_0$, in which case only austenite is stable. When the temperature is much lower than
$\theta_0$, $\phi$ becomes non-convex and  has two local minima associated with two
martensite variants, which are the only stable variants. If the temperature is around
$\theta_0$,  the free energy function has totally three local minima, two of which are
symmetric and associated with  the martensitic phases and the remaining one is associated
with the austenitic phase. In this case both martensite and austenite phases could co-exist
in the system. By substituting the above free energy function into the conservation laws for
momentum and energy, and using Fourier's heat flux definition
\begin{equation}
q=-k\theta_{x}
\label{eq2-6}
\end{equation}
with $k>0$ being the heat conductivity of the material, the governing equations for 2D SMA
patches can be written in the following form:
\begin{equation}
\label{eq2-7}
\begin{array}{l} \displaystyle
\rho \frac{\partial^2 u_1}{\partial t^2} = \frac{\sqrt{2}}{2} \frac{\partial}
{\partial x} \left ( a_1 e_1 + a_2 (\theta-\theta_0) e_2 - a_4 e_2^3 + a_6
e_2^5\right )+ \frac{\partial }{\partial y}
\left ( \frac{1}{2}  a_3 e_3 \right ) +f_1,
\\[10pt] \displaystyle
\rho \frac{\partial^2 u_2}{\partial t^2} = \frac{\partial }{\partial x}
\left ( \frac{1}{2} a_3 e_3 \right) + \frac{\sqrt{2}}{2} \frac{\partial}
{\partial y} \left (  a_1 e_1 - a_2 (\theta - \theta_0) e_2 + e_4 e_2^3 - a_6
e_2^5\right )+f_2 ,
\\[10pt]  \displaystyle
c_v \frac{\partial \theta}{\partial t} = k \left (\frac{\partial^2
\theta}{\partial x^2} + \frac{\partial^2 \theta}{\partial y^2} \right )
+  a_2 \theta e_2 \frac{\partial e_2}{\partial t} + g.
\end{array}
\end{equation}
As always, we complete system (\ref{eq2-7}) by appropriate initial and boundary
conditions which are problem specific (see Sections 4 and 5).  As discussed before
in \cite{Matus2003,Wang2004}, the 2D model given by
Eq.\ref{eq2-7} can be reduced to the Falk model in the 1D case
\begin{equation}
\label{eq2-8}
\begin{array}{l}  \displaystyle
\rho\frac{\partial^{2} u}{\partial t^{2}}=\frac{\partial}{\partial
x}\left(k_{1}\left(\theta-\theta_{1}\right)\frac{\partial u}{\partial
x}-k_{2}(\frac{\partial u}{\partial x})^{3}+k_{3}(\frac{\partial u}{\partial
x})^{5}\right)+F,
   \\[10pt] \displaystyle
c_{v}\frac{\partial\theta}{\partial
t}=k\frac{\partial^{2}\theta}{\partial x^{2}}+k_{1}\theta \frac{\partial
u}{\partial x}\frac{\partial v}{\partial t}+G,
\end{array}
\end{equation}
\noindent where $k_{1}$, $k_{2}$, $k_{3}$, $c_{v}$ and $k$ are
re-normalized material-specific constants, $\theta_1$ is the reference
temperature for 1D martensitic transformations,  and $F$ and $G$ are distributed
mechanical and thermal loadings.

In the subsequent sections, the above models are applied to the description of the first
order martensitic transformations. While such transformations are reasonably well documented
for the 1D case, only few results are known for the 2D case.  In what follows, we develop a
FVM to simulate the dynamics described  by the models (\ref{eq2-7}) and (\ref{eq2-8}) and
apply it in both 1D and 2D cases.


\section{Numerical Algorithm}

The systems (\ref{eq2-7}) and (\ref{eq2-8}) are analyzed numerically with the FVM implemented
here with the help of the DAE approach.  For the 1D case, the FVM method yields the same
result as the conservative scheme already discussed in \cite{Matus2004}.  However, the
approach developed here is generalized in a straightforward manner to a higher dimensional
case and we demonstrate its applicability by a numerical example in the case of two spatial
dimensions. First, we note that
 it is convenient to replace the original
model (\ref{eq2-7})  by a system of
equivalent differential-algebraic equations as it was proposed earlier in \cite{Melnik2000,Melnik2002}:
\begin{equation}
\label{eq3-1}
\begin{array}{l}
\displaystyle
\frac{\partial e_1}{\partial t} = \frac{\sqrt 2}{2} \left(  \frac{\partial
v_1}{\partial x}+\frac{\partial v_2}{\partial y} \right),
\quad
\frac{\partial e_2}{\partial t} = \frac{\sqrt 2}{2} \left(  \frac{\partial
v_1}{\partial x} - \frac{\partial v_2}{\partial y} \right),
\\[10pt]  \displaystyle
\rho \frac{\partial v_1}{\partial t}= \frac{\partial\sigma_{11}}{\partial
x}+\frac{\partial\sigma_{12}}{\partial y}+f_{x},
\quad
\rho \frac{\partial v_2}{\partial t}=\frac{\partial\sigma_{12}}{\partial
x}+\frac{\partial\sigma_{22}}{\partial y}+f_{y},
\\[10pt] \displaystyle
c_{v}\frac{\partial\theta}{\partial t}=k\left(\frac{\partial^{2}\theta}{\partial
x^{2}}+\frac{\partial^{2}\theta}{\partial y^{2}}\right)+a_2\theta
e_{2}\frac{\partial e_{2}}{\partial t}+g,
\\[10pt]\displaystyle
\sigma_{11}=\frac{\sqrt{2}}{2}(a_1e_{1}+a_2\left(\theta-\theta_{0}\right)e_{2}-
a_4 e_{2}^{3}+ a_6e_{2}^{5}),
\\[10pt] \displaystyle
\sigma_{12}=\sigma_{21}=\frac{1}{2}a_3e_{3},
\\[10pt] \displaystyle
\sigma_{22}=\frac{\sqrt{2}}{2}(a_1 e_{1}-
a_2 \left(\theta-\theta_{0}\right)e_{2} + a_4 e_{2}^{3} - a_6e_{2}^{5}).
\end{array}
\end{equation}
This system is solved  numerically together with the compatibility relation written below in
terms of strains:
\begin{equation}
\label{eq3-2}
\frac{\partial^{2}e_{1}}{\partial x_{1}^{2}}+\frac{\partial^{2}e_{1}}
{\partial x_{2}^{2}}-\sqrt{8}\frac{\partial^{2}e_{3}}{\partial x_{1}\partial
x_{2}}-\frac{\partial^{2}e_{2}}{\partial
x_{1}^{2}}+\frac{\partial^{2}e_{2}}{\partial x_{1}^{2}}=0 .
\end{equation}
There are  eight variables in total that the problem needs to be solved for in this 2D case
and there are eight equations. The equations for strains, velocities and temperature are all
differential equations, complemented by stress-strain relationships which are treated as
algebraic.

In what follows, we highlight the key elements of our numerical procedure based on the FVM
implemented with the help of the DAE approach.  First, all equations in the system
(\ref{eq3-1}) are discretized on a staggered grid represented schematically in Fig 1.
Assuming that the entire computational domain is a rectangle with an area of $ L_x \times L_y
$ ${\rm cm}^2$, we define the spatial integer grid points $(x_{i},y_j)$ and the spatial flux
points $ (\bar{x}_{i}, \bar y_j )$ as follows:
\begin{equation}
\label{eq3-3}
\begin{array}{l} \displaystyle
x_{i}=ih_x,\quad, i=0,1,2,\cdots,M,\qquad\bar{x}_{i}=(i-\frac{1}{2})h_x,\quad,
i=1,2,\cdots,M \\[10pt] \displaystyle
y_{j}=jh_y,\quad, j=0,1,2,\cdots,N,\qquad\bar{y}_{j}=(j-\frac{1}{2})h_y,\quad,
j=1,2,\cdots,N
\end{array}
\end{equation}
where $M$ and $N$ are the number of grid points such that $M\times h_x=L_x$ and $N\times
h_y=L_y$, respectively. The $(i,j)$th control volume for the velocities is
$[\bar{x}_{i},\bar{x}_{i+1} ] \times [\bar{y}_{j},\bar{y}_{j+1} ]$, as sketched by the
rectangular tiled mosaic area in Fig.1, including the upper right part overlapped with the
hatched area. The variables, defined in this control volume, that will be  differentiated are
marked by a top bar, for instance $\bar{v}_1(i,j)$.  The control volume for the strains $e_1$
and $e_2$, temperature $\theta$, and stresses $\sigma_{11}$, $\sigma_{12}$, and $\sigma_{22}$
is given by  $[x_{i},x_{i+1} ] \times [y_{j},y_{j+1} ]$, represented by  the rectangular
hatched area in Fig. 1. We refer to these variables, defined in this control volume, without
a top bar, for instance $e_1(i,j)$ for $e_1$, etc.

By integrating all the differential equations over their own control volumes
and
 assuming that all the
unknowns are linear in each single control volume while being continuous and piecewise linear
in the entire computational domain, the five partial differential equations are reduced to a
system of ordinary differential equations. The remaining three algebraic equations of the
original system are discretized directly on the grid. The result is the following system:
\begin{equation}
\label{eq3-4}\begin{array}{l}
 \displaystyle
\frac{d e_1(i,j)}{dt}=(I_y D_x \bar v_1(i,j) +
          I_x D_y \bar v_2(i,j))/ \sqrt{2},
     \\[10pt] \displaystyle
\frac{d e_2(i,j)}{dt}=( I_y D_x \bar v_1(i,j) - I_x D_y
       \bar v_2(i,j))/ \sqrt{2},
     \\[10pt] \displaystyle
\rho \frac{d \bar v_1(i,j)}{d t} =
    I_y D_x \sigma_{11}(i-1,j-1) + I_x D_y \sigma_{12}(i-1,j-1) + f_{1},
       \\[10pt] \displaystyle
\rho \frac{d \bar v_2(i,j)}{d t} =
     I_yD_x \sigma_{12}(i-1,j-1) +I_x D_y \sigma_{22}(i-1,j-1) + f_{2},
        \\[10pt] \displaystyle
c_v \frac{ d\theta(i,j)}{dt} = k (\triangle \theta(i,j))
+ \frac{\sqrt{a_2}}{2} \theta(i,j) e_2(i,j) \frac{d e_2}{dt} + g,
      \\[10pt] \displaystyle
\sigma_{11}(i,j) =  \frac{\sqrt{2}}{2}(a_{1}e_{1}(i,j)+a_{2}(\theta(i,j)-
\theta_{0})e_{2}(i,j)-\frac{a_4}{4}g_{1}(i,j)+\frac{a_6}{6}g_{2}(i,j)) ,
      \\[10pt] \displaystyle
\sigma_{12}(i,j)=\sigma_{21}(i,j)= \frac{1}{2}(a_{3}e_{3}(i,j) ),
       \\[10pt] \displaystyle
\sigma_{22}(i,j) = \frac{\sqrt{2}}{2}( a_{1}e_{1}(i,j)- a_{2}(\theta(i,j)
-\theta_{0})e_{2}(i,j)+\frac{a_4}{4}g_{1}(i,j)-\frac{a_6}{6}g_{2}(i,j)).
\end{array}
\end{equation}
where $D_x$ and $D_y$ are the discrete difference operators in the $x$ and $y$ directions,
respectively, while $I_x$ and $I_y$ are the discrete interpolation operator in the $x$ and
$y$ directions, and $\triangle$ is the discrete Laplace operator. For example, for the
simplest case of the first order accurate scheme, the operators $D_x$ and $I_y$ could be
written as follows
\begin{equation}
  D_x \bar v_1(i,j) = ( \bar v_1(i,j+1) - \bar v_1(i,j) ) / h_x , \quad
  I_y \bar v_1(i,j) = ( \bar v_1(i,j)  + \bar v_1(i+1,j) ) / 2,
\end{equation}
with similar representations for the second order accurate schemes.
Moving to the time discretization procedure, it is convenient to re-write
system (\ref{eq3-4}) in the following vector-matrix form:
\begin{eqnarray}
\label{eq3-5}
{\bf{A}}\frac{{d{\bf{U}}}}{{dt}} + {\bf{H}}\left(
{t,{\bf{X}},{\bf{U}}} \right) = {\bf{0}}
\end{eqnarray}
with matrix $A={\rm diag}(a_1, a_2,...,a_N)$ having entries ``one'' for differential and
``zero'' for algebraic equations for stress-strain relationships, and vector-function ${\bf
H}$ defined by the right hand side parts of (\ref{eq3-4}). This (stiff) system is solved with
respect to the vector of unknowns ${\bf U}$ that have $6 \times m_x \times m_y + 2 \times
(m_x+1) \times (m_y+1)$ components by using the second order backward differentiation formula
(BDF) \cite{Hairer1996}:
\begin{equation}
\label{eq3-6}
{\rm {\bf A}}\left( {\frac{3}{2}
{\rm {\bf U}}^n - 2{\rm {\bf U}}^{n - 1} +\frac{1}{2}{\rm {\bf U}}^{n - 2}}
\right) + \Delta t{\rm {\bf H}}\left( {t_n,{\rm {\bf X}},{\rm {\bf U}}^n}
\right) = 0
\end{equation}
where $n$ denotes the current time layer.

This spatio-temporal discretization is applied to the analysis of phase
transformations with the following modification. In order to improve
convergence properties of the scheme, we employ a
relaxation process connecting two consecutive time layers via  a relaxation factor $\omega$ as follows:
\begin{equation}
\label{eq3-7}
y(i,j) =(1-\omega) \times y(i,j)^n + \omega \times y(i,j) ^ {n+1},
\end{equation}
where the variable $y$ could be any of the following:  $e_1(i,j)$, $e_2(i,j)$, $v_1(i,j)$,
$v_2(i,j)$, or $\theta(i,j)$. Note that in the general case the relaxation factors need not
be the same for all the variables. In the present paper, all the numerical results have been
obtained using (\ref{eq3-7}) with all the relaxation factors set to $0.85$.

We note that nonlinear terms in the model are averaged in the Steklov sense \cite{Matus2004},
so that for nonlinear function $f(e_2)$ (in particular,  for $e_2^3$ and $e_2^5$), averaged
in the interval $[e_2^{n}, e_2^{n+1}]$, we have
\begin{equation}
\label{eq3-8}
g\left( e_2^{n}, e_2^{n+1}\right ) =
\frac{1}{e_2^{n+1} - e_2^{n}} \int_{e_2^{n}}^{e_2^{n+1}}
f(e_2)d e_2.
 \end{equation}
Applying this idea to  $e_2^3$ and $e_2^5$, we have:
\begin{equation}
\label{eq3-9}
\begin{array}{c}
\displaystyle
   g_1 \left( i,j \right)    =
\frac{\left(e_2^{n+1}\right)^{4}-(e_2^{n})^{4}}{e_2^{n+1}-e_2^n}=
\frac{1}{4}\sum_{k=0}^{3} (e_2^{n+1})^{3-k} (e_2^n)^{k},
 \\[10pt] \displaystyle
g_2\left(  i,j \right ) =
\frac{(e_2^{n+1})^6-(e_2^{n})^6}{e_2^{n+1}-e_2^n} =
\frac{1}{6} \sum_{k=0}^5 (e_2^{n+1})^{5-k} (e_2^n)^{k}
.\end{array}
\end{equation}
where  $e_2^{n}$ and $e_2^{n+1}$ stands for $e_2(i,j)^n$ and  $e_2(i,j)^{n+1}$, respectively.

Finally, we note that in our FVM implementation the nonlinear coupling term in the energy
balance equation is regarded as a time-dependent source term. In the $(i,j)$th control volume
for the discretization of $\theta$, we approximate that term as follows:
\begin{equation}
\label{eq3-10}
\int_{x_{i}}^{x_{i+1}}\int_{y_{j}}^{y_{j+1}}\left(k_{1}\theta e_2
\frac{\partial e_2}{\partial t}\right) dxdy \approx
\theta(i,j) e_2(i,j)\frac{d e_2(i,j)}{d t}.
\end{equation}
As seen from (\ref{eq3-6}), we use an implicit time integrator based on the
BDF. At each time step we apply the bi-conjugate gradient method to
solve the resultant system of algebraic equations with the Jacobian matrix updated on
each iteration.


\section{Dynamics of SMA Rods and Strips}

We first consider a situation where the deformation of a 2D SMA sample in the $x_1$ direction
substantially exceeds the deformation in the other direction, so that the deformation in the
$x_2$ direction can be neglected and the sample can be treated as a SMA long strip or simply
as a rod. Introducing formally  $\eps= \partial{u}/\partial x$ and  $ v=
\partial{u}/\partial t$, system (\ref{eq2-8}) can be recast in the following form:
\begin{equation}
\label{eq4-1}
\begin{array}{c}
\displaystyle
\frac{{\partial \epsilon }}{{\partial t}} = \frac{{\partial v}}{{\partial x}},
   \quad  {\rho \frac{{\partial v}}{{\partial t}} = \frac{{\partial
s}}{{\partial x}} + F},
  \\[10pt]   \displaystyle
s =  {k_1 \left( {\theta  - \theta _1 } \right)\epsilon - k_2 \epsilon ^3  + k_3
\epsilon ^5 },
   \\[10pt]   \displaystyle
 c_v \frac{{\partial \theta }}{{\partial t}} = k\frac{{\partial ^2 \theta
}}{{\partial x^2 }} + k_1 \theta \epsilon \frac{{\partial v}}{{\partial x}} + G,
\end{array}
\end{equation}
where $\epsilon$ is strain and $s$ is stress.

The numerical procedure described in Section 3 is applied here to the solution of system
(\ref{eq4-1}). It is aimed at the analysis of martensitic transformations in the SMA rod,
including  hysteresis effects during the transformations. Computational experiments reported
in this section were performed for a $Au_{23}Cu_{30}Zn_{47}$ rod with a length of $L=1$ ${\rm
cm}$ and all parameter values found in \cite{Falk1980,Melnik2001,Niezgodka1991}, in
particular:
\begin{displaymath}
\label{eq4-3}
\begin{array}{c}
\displaystyle
 k_{1}=480\, g/ms^{2}cmK,\qquad
 k_{2}=6\times10^{6}g/ms^{2}cmK,\qquad k_{3}=4.5\times10^{8}g/ms^{2}cmK,
\\[10pt]\displaystyle
\theta_{1}=208K,\quad  \rho=11.1g/cm{}^{3},\quad
C_{v}=3.1274g/ms^{2}cmK, \quad k=1.9\times10^{-2}cmg/ms^{3}K.
\nonumber
\end{array}
\end{displaymath}

The boundary conditions for $u$ and $\theta$ for all the numerical experiments reported in
this section are:
\begin{equation}
\label{eq4-4}
\begin{array}{c} \displaystyle
u(0,t)= u_L(t), \quad u(L,t)= u_R(t),qquad \frac{\partial
\theta}{\partial x}(0,t)= \theta_L(t), \quad  \frac{\partial
\theta}{\partial x}(L,t)= \theta_R(t)
\end{array}
\end{equation}
with given functions $u_i(t)$ and $\theta_i(t)$, $i=L, R$ and corresponding conditions for
the velocities.

In the numerical experiments reported below, we used only $9$ nodes  for the velocity
discretization (and 8, excluding boundaries, for the rest of variables). The time stepsize in
all the experiments was set to $\tau=1. \times 10^{-4}$. All the simulations were performed
for the time period  $[0 , 24] $ which spans two periods of the loading cycle.


\subsection{Mechanically Induced Transformations and Hysteresis}

The first numerical experiment deals with the case of mechanical loading in the
low-temperature regime. The initial conditions for this computational experiment are defined
by the following configuration of martensites (\cite{Klein1995,Melnik2001,Melnik2003}):
\begin{equation}
\label{eq4-13}
\theta(x,0)=220,\qquad u_{0}=\left\{ \begin{array}{l}
0.11869x,\qquad  \qquad 0\leq x\leq0.25\\
0.11869(0.5-x), \quad  0.25\leq x\leq0.75\\
0.11869(x-1),\qquad   0.75\leq x\leq1\end{array}\right.,v^{0}=u^{1}=0
\end{equation}
with  the time varying distributed mechanical loading defined as
\begin{equation}
\label{eq4-14}
F=7000\sin^{3}\left(\frac{\pi t}{2}\right)\,\, g/\left(ms^{2}cm^{2}\right),\,\,\,\,
G=0.
\end{equation}

Under the given distributed mechanical loading, the SMA rod is expected to switch between
different combinations of the martensite variants, and a hysteresis loop must be observed
similar to those reported for ferroelastic materials at low temperature.  In Fig. 2 we
present simulation results for this case.  The mechanical hysteresis is obtained by plotting
displacement $u$ as a function of $F$ at $x=3/8$cm (the upper right plot). The time-varying
mechanical loading for this case is plotted in the upper left plot.
 The simulated strain and
the displacement distribution are also plotted as functions of time and space (lower plots).
The combination of martensitic variants is changing with time-dependent mechanical loading
 and no stable austenite is observed at this low temperature.

Our next goal is to analyze the behavior of the same SMA rod under a medium temperature where
both martensite and austenite phases may co-exist. The following initial conditions will
allow us to start from the austenitic phase:
\begin{equation}
\label{eq4-15} \theta(x,0)=250,\,\, u_{0}=0,\,\,\, v^{0}=u^{1}=0.
\end{equation}
The boundary conditions as well as mechanical and thermal loadings in this case are kept
identical the previous experiment. In this case, the free energy function has three minima
that correspond to two martensites and one austenite.  The numerical results for this case
are presented in the left column of Fig.3. It is observed that when the applied  loading
exceeds a certain value, the austenite is transformed to a combination of martensitic
variants. The reverse transformation is taken place when the loading changes its sign. In
contrast to the results presented in Fig. 2, we observe that the wide hysteresis loop,
typical for the low temperature case, disappears.

If we increase the initial temperature further to $\theta(x,0)=300$, the free energy function
becomes convex and has only one minimum associated with the austenite phase. During the
entire loading cycle, no martensite is expected under these thermal conditions. The dynamics
of the SMA rod in this case exhibits nonlinear thermomechanical behaviour without phase
transformations. This is confirmed by the numerical results presented in the right column of
Fig.3.


\subsection{Thermally induced Phase Transformations and Hysteresis}

Thermally induced martensitic phase transformations and thermal hysteresis in SMA rods can be
analyzed with the same model under time-dependent thermal loading conditions. Indeed, let us
choose the initial conditions as follows:
\[
\theta(x,0)=230,\,\, u_{0}=\left\{ \begin{array}{l}
0.11869x,\qquad \qquad  0\leq x\leq0.5\\
0.11869(1-x),\quad 0\leq x\leq0.5\end{array}\right.,v^{0}=u^{1}=0\]

The boundary conditions remain the same as in the previous computational experiment, but the
loadings conditions now become:
\[
G=600 \sin \left(\frac{\pi t}{6}\right)\,\, g/\left(ms^{3}cm\right),\,\,\, F=500\,
g/\left(ms^{2}cm^{2}\right).  \]

Numerical results for this case are presented in Fig.4. Analyzing strain and displacement
distributions, we observe that the combination of martensitic variants is transformed into
the austenite phase when the temperature exceeds a certain value. The reverse process is
taken place when the temperature decreases, passing the critical threshold. Note that due to
the presence of thermal hysteresis, the critical temperature value for the
martensite-to-austenite transformation is different from that of the austenite-to-martensite
transformation. A schematic representation of the observed thermal hysteresis is given in the
lower right part of Fig. 4 where we presented the temperature at $x=3/8$  as a function of
strain at the same spatial point.


\section{Dynamics of SMA Patches}

The situation becomes more involved for 2D structures. Experimental, let alone numerical,
results for this situation are scarce \cite{Wang2004}.  In order to apply the FVM to the 2D
model discussed in Section 2, we chose the same material as before, assuming that
$a_{2}=k_{1}, a_{3}=k_{2},\, a_{4}=k_{3}, a_{1}=k_{1},\, a_{3}=2k_{1}$ and therefore
effectively linking parameters in model (\ref{eq2-7}) and model (\ref{eq2-8}).

\subsection{Nonlinear Thermomechanical Behavior}

The first numerical experiment on a SMA patch is aimed at the analysis of the dynamical
thermo-mechanical response of the patch to a varying distributed mechanical loading, too
small to induce any phase transformations. The initial temperature of the patch is set to
$250^o$ while all other variables are set initially to zero. Conditions at the boundaries
are:
 \begin{equation}
 \begin{array}{l}  \displaystyle
 \frac{\partial \theta}{\partial x}=0,  \quad \frac{\partial u_2}{\partial x}=0,
 \quad u_1=0, \quad \textrm{on left and right boundaries},  \\[10pt]
 \displaystyle
  \frac{\partial \theta}{\partial y}=0,  \quad \frac{\partial u_1}{\partial
y}=0,  \quad u_2=0, \quad \textrm{on top and bottom boundaries}.
 \end{array}
 \label{eq11}
 \end{equation}
Similarly, the mechanical boundary conditions are enforced in terms of velocity components.
The loading conditions in this experiment are:
\begin{displaymath}
f_1 = 200 \sin ( \pi t /6 ) {\textrm{g}}/(\textrm{ms}^{2}\textrm{cm}^{2}), \quad
f_2 = 200 \sin ( \pi t /40 ) {\textrm{g}}/(\textrm{ms}^{2}\textrm{cm}^{2}).
\end{displaymath}
The time span for this simulation, $[0,24]$, covers two periods of loading. The time stepsize
is set to $1\times 10^{-4}$. We take $15$ nodes used in each direction. The dimensions of the
SMA patch are taken as $1 \times 0.4$ ${\rm cm}^2$.

The variations in the displacements $u_1$, $u_2$ , deviatoric strain $e_2$, and the
temperature $\theta$ along the line $y=0.2\rm{cm}$ (the central horizontal line) as functions
of time are presented in Fig.5.  These simulations show that both thermal and mechanical
fields are driven periodically by the distributed mechanical loading. Under such a small
loading, the SMA patch behaves just like a conventional thermoelastic material. Observed
oscillations are due to nonlinear thermomechanical coupling, but no phase transformations are
observed in this case.

\subsection{Phase Transformations in SMA Patches}

Our aim in this section is to analyze spatio-temporal patterns of martensitic transformations
in a 2D SMA patch. The SMA patch, used in this computational experiment, is made of the same
material as before. The patch is assumed square in shape with dimensions $1 \times 1$ ${\rm
cm}^2$. The initial temperature distribution is set to $\theta^0=240^o $, and all other
variables are initially set to zero. The boundary conditions are homogeneous and
 \begin{equation}
  \frac{\partial \theta}{\partial  n}=0,  \quad u_2 = u_1 =0,
  \quad \textrm{on all the four boundaries,}
  \end{equation}
where n is the unit normal vector. We apply the following loading to the sample, specified
below for one period:
\begin{eqnarray}
  f_1 = f_2 = 6000 \left \{
 \begin{array}{llll}
 \sin (\pi t/3), & 0 \leq t \leq 4, \\
  0 ,\qquad  &  4 \leq t \leq 6, \\
 \sin(\pi (t-2))/3), &  6 \leq t \leq 10, \\
  0 ,\qquad  &  10 \leq t \leq 12. \\
 \end{array}
 \right.
\end{eqnarray}
The numerical results for this case are presented in Fig.6 where the values of "y" are taken
in the middle of the sample. The spatio-temporal plot of the order parameter $e_2$
demonstrates a periodicity pattern in the observed phase transformations due to periodicity
of the loading. It is observed also that the temperature oscillates synchronously with the
mechanical field variables due to the thermo-mechanical coupling.

As we mentioned earlier, there are two martensitic variants in the square-to-rectangular
transformations. The following analysis proves to be useful in validating the results of
computational experiments. Assuming the temperature difference $d\theta  = \theta-\theta_0$,
one can easily calculate the deviatoric strain that corresponds to the austenite and
martensite variants by minimizing the Landau free energy functional. In particular from the
condition $\partial{F_l}/\partial{e_2}=0$ we get:
  \begin{displaymath}
  e_2  = 0 ; \quad  e_2^2 = \frac{a_4\pm \sqrt{a_4^2
      -4a_2d\theta a_6}}{2a_6}.
 \end{displaymath}
The value $e_2=0$ corresponds to the austenitic phase. If we denote $(a_4 +
\sqrt{a_4^2-4a_2d\theta a_6})/2a_6$ by $e_m$, then $e_{2+} = +\sqrt{e_m}$ or $e_{2-} =
-\sqrt{e_m}$ are the strains that correspond to the two martensite variants. We call them
martensite plus and martensite minus, respectively. If we take $d\theta = 42^o $, then  for
the material considered here we can estimate that $e_{2+} = 0.12$ and $e_{2-} = 0.12$. This
provides a fairly good estimate for the 1D case. However, as was pointed out in
\cite{Jacobs2000,Lookman2003}, for the 2D case such an estimate can be adequate only
 in homogenous cases.  Although the quality of this estimate is dependent on the boundary
 conditions for a specific problem, this estimate proves to be a reasonable initial approximation to
 the deviatoric strain.

 In Fig. 7 we present two snapshots (at $t=2$ and $t=8$) of the spatial distributions of $e_2$ and $\theta$.
 It is observed that when the mechanical loading achieves its (positive) maximum,
  the SMA patch is divided into two sub-domains determined by the
deviatoric strain, as seen from the $e_2$ plot at $t=2$. In the upper-left triangular-shape
area, the simulated deviatoric strain corresponds to the martensite plus, while on the
opposite side, the deviatoric strain corresponds to the martensite minus.  At $t=8$, when the
mechanical loading changes its sign to the opposite, the martensitic transformation is
observed again, but now in the reverse direction.  The second period of loading confirms
these observations.

\section{Conclusion}

In this paper, we developed a finite volume methodology for the analysis of nonlinear coupled
thermomechanical problems, focusing on the dynamics of SMA rods and patches. Both
mechanically and thermally induced phase transformations, as well as hysteresis effects, in
one-dimensional structures are successfully simulated. While these results can be obtained
with the recently developed conservative difference schemes, their generalization to higher
dimensional cases is not trivial. In this paper, we also highlighted the application of the
developed FVM to the 2D problems focusing on  square-to-rectangular transformations in SMA
materials demonstrating practical capabilities of the developed methodology.



\newpage

\begin{figure}
\includegraphics[ scale=0.6]{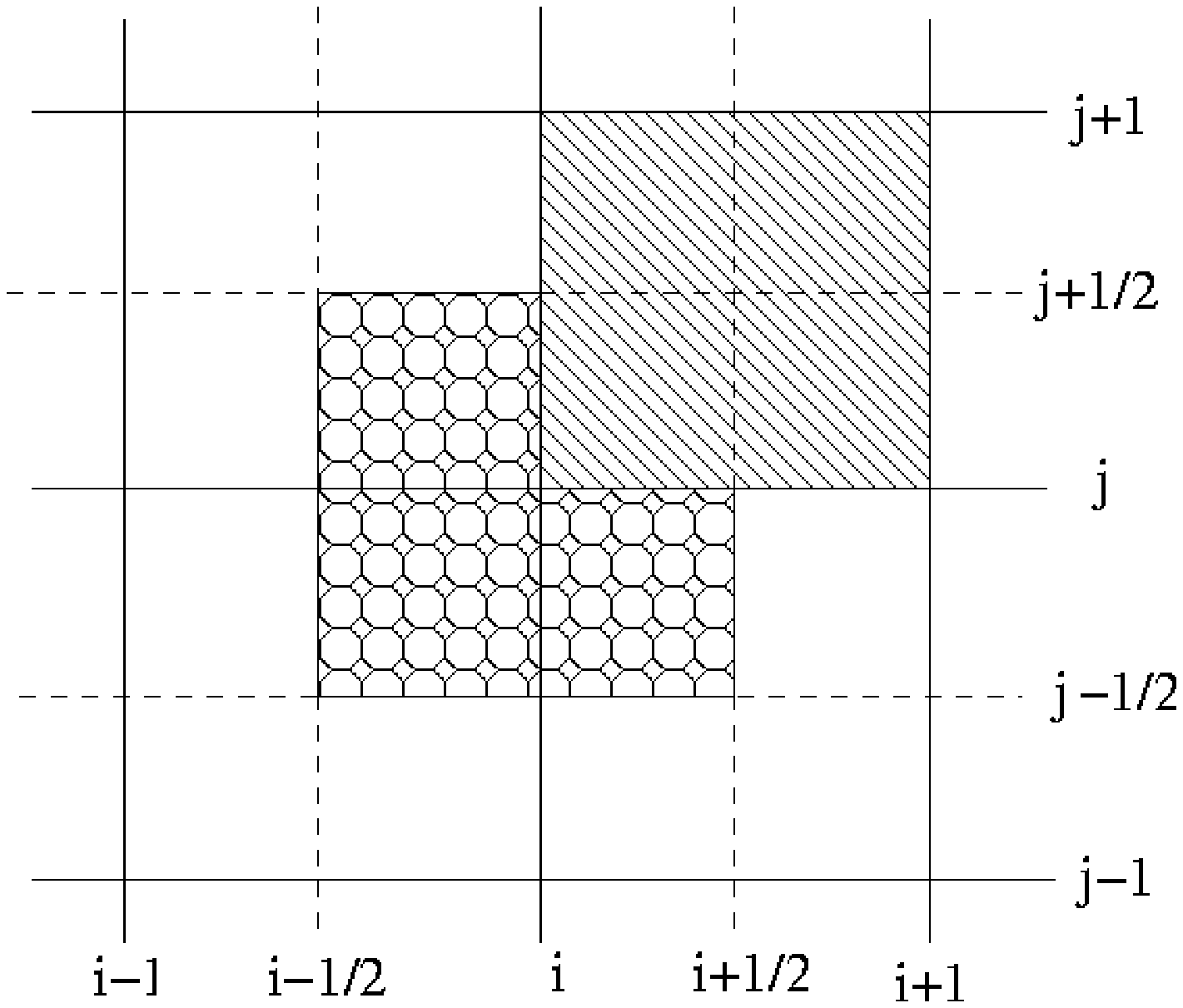}
\caption{Staggered grid for space discretization using finite volume methods.}
\end{figure}


\newpage

\begin{figure}
  \includegraphics[scale=0.4]{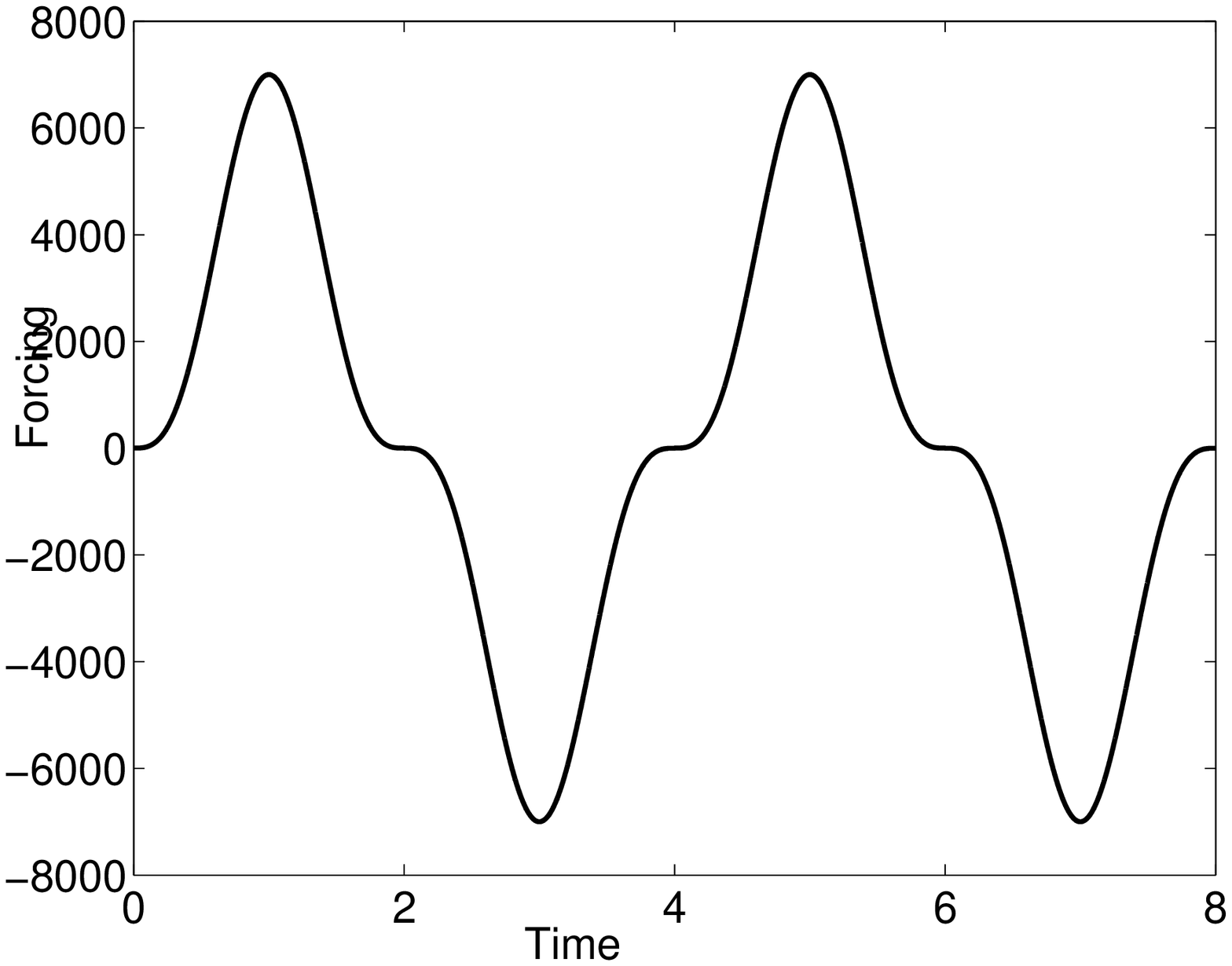}
  \includegraphics[scale=0.4]{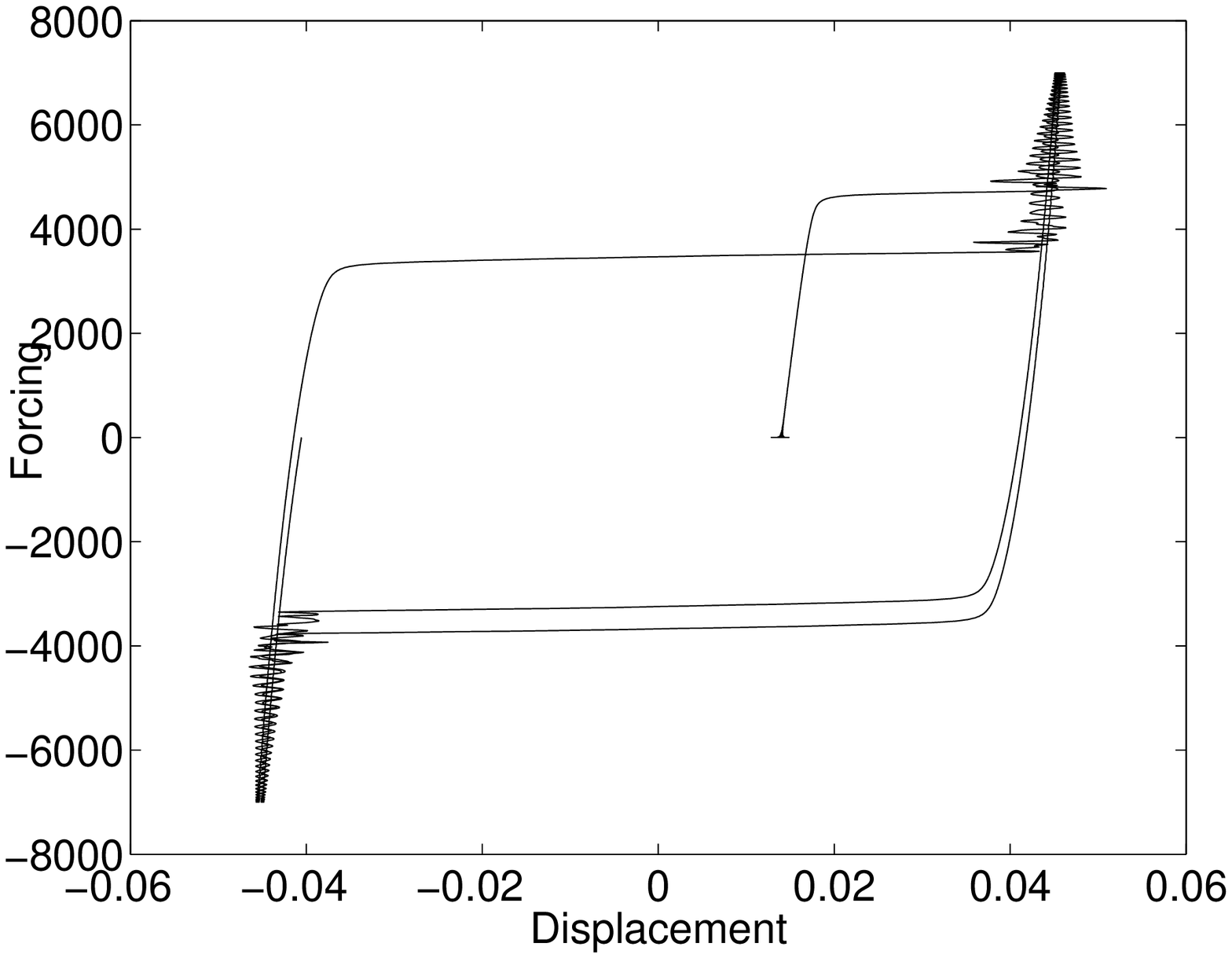}  \\
  \includegraphics[scale=0.4]{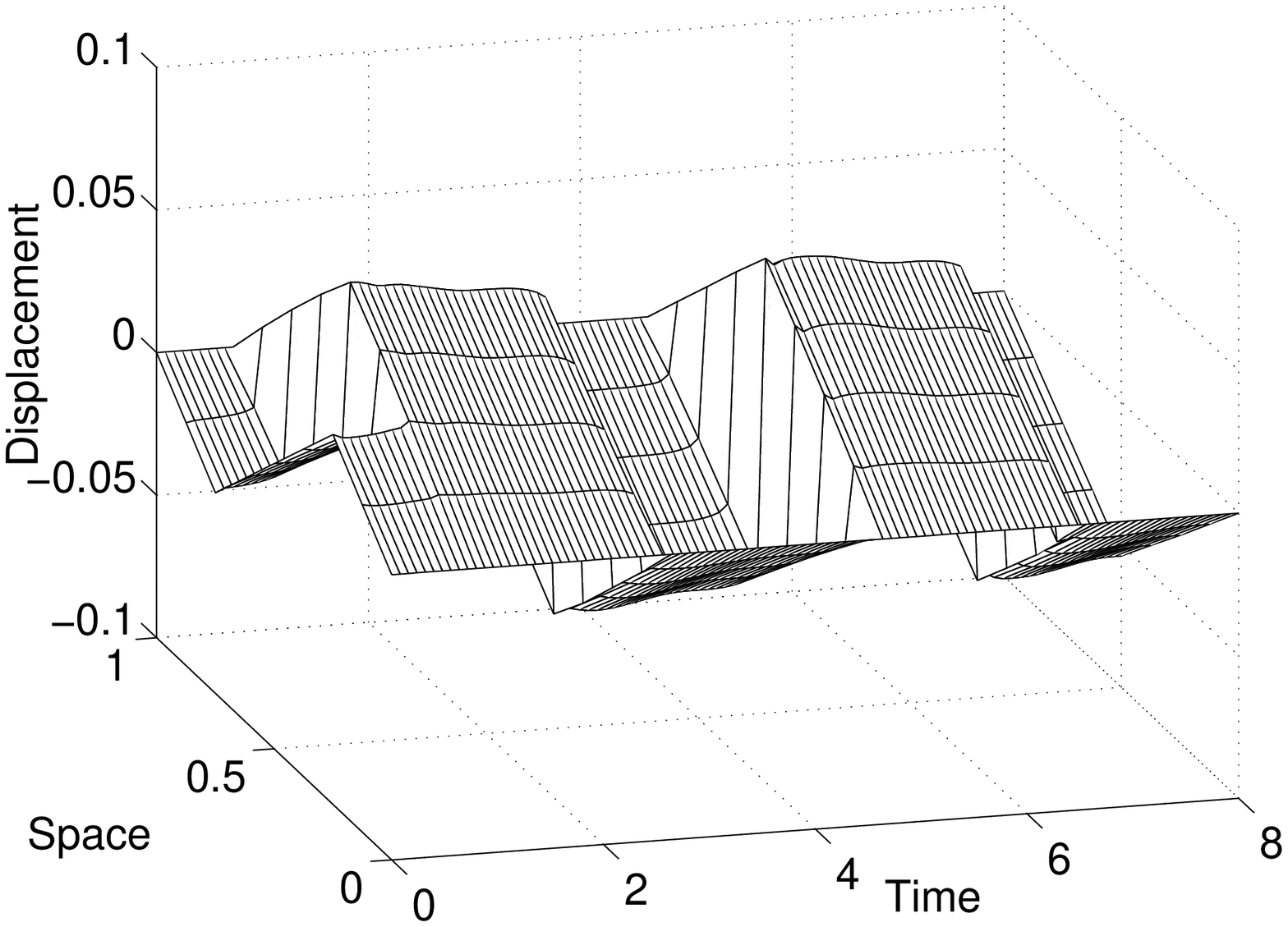}
  \includegraphics[scale=0.4]{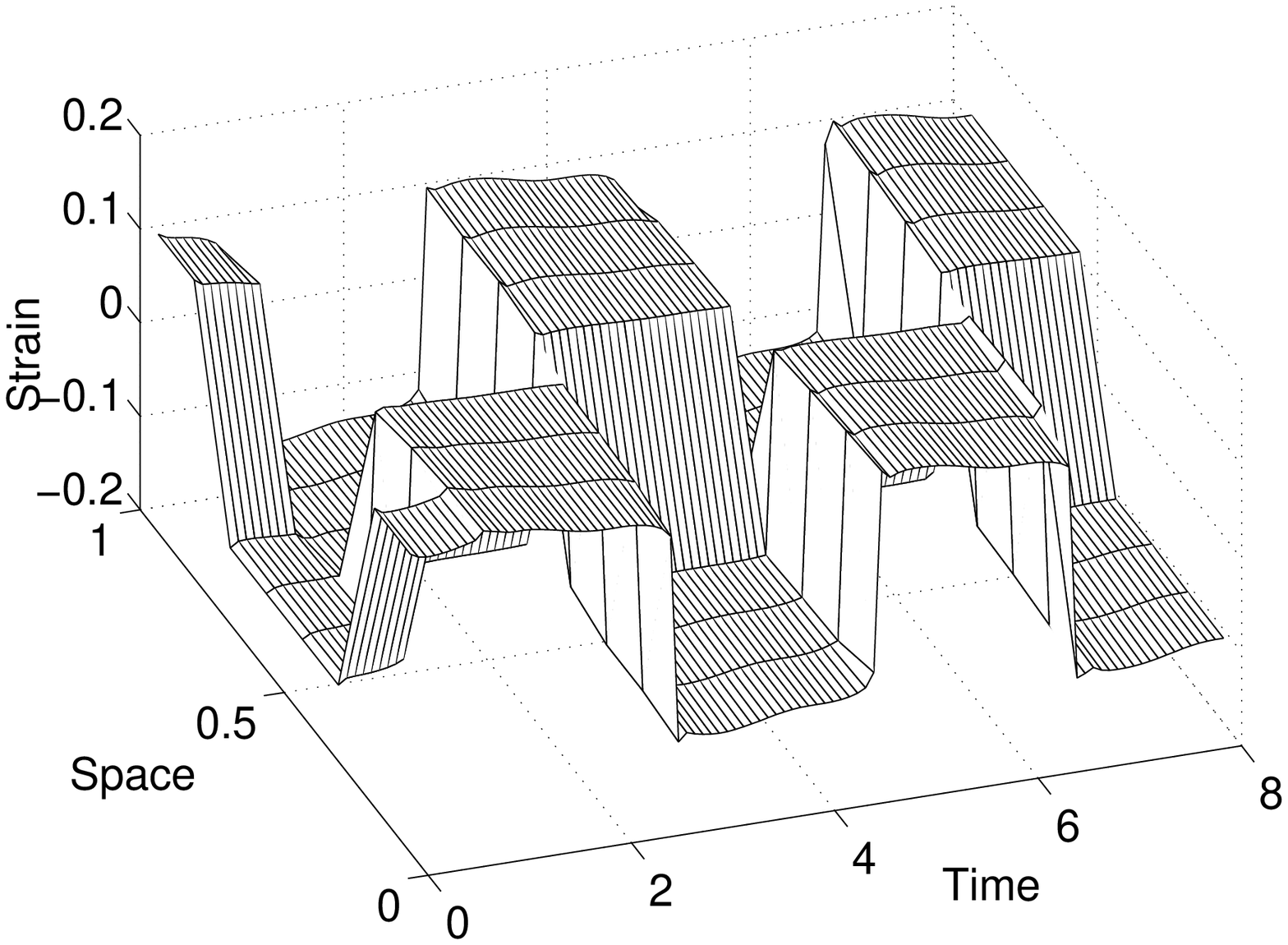}  \\
  \caption{Mechanical induced phase transformation and mechanical hysteresis
in SMA rod.}
\end{figure}


\newpage

\begin{figure}
  \includegraphics[scale=0.4]{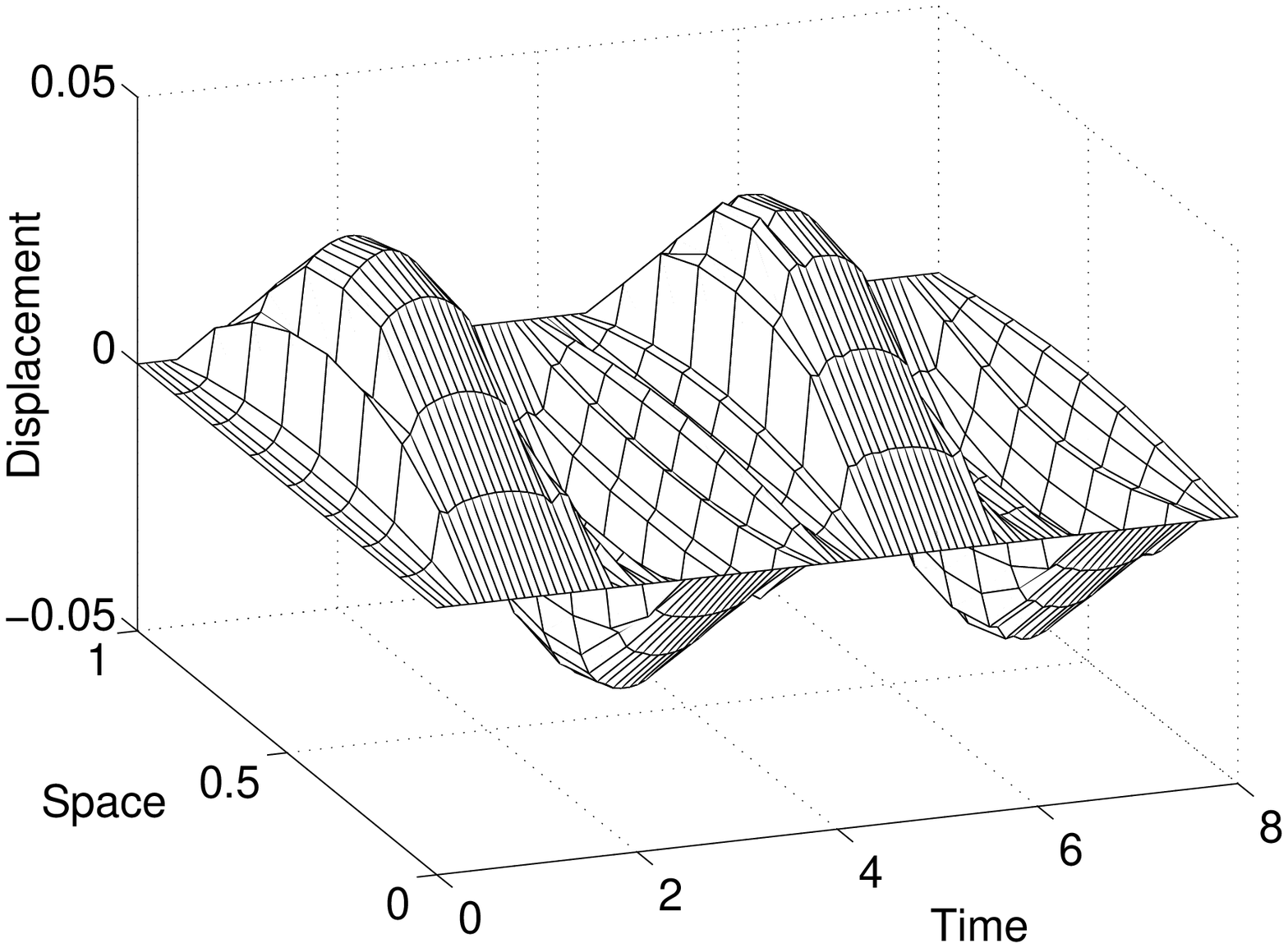}
  \includegraphics[scale=0.4]{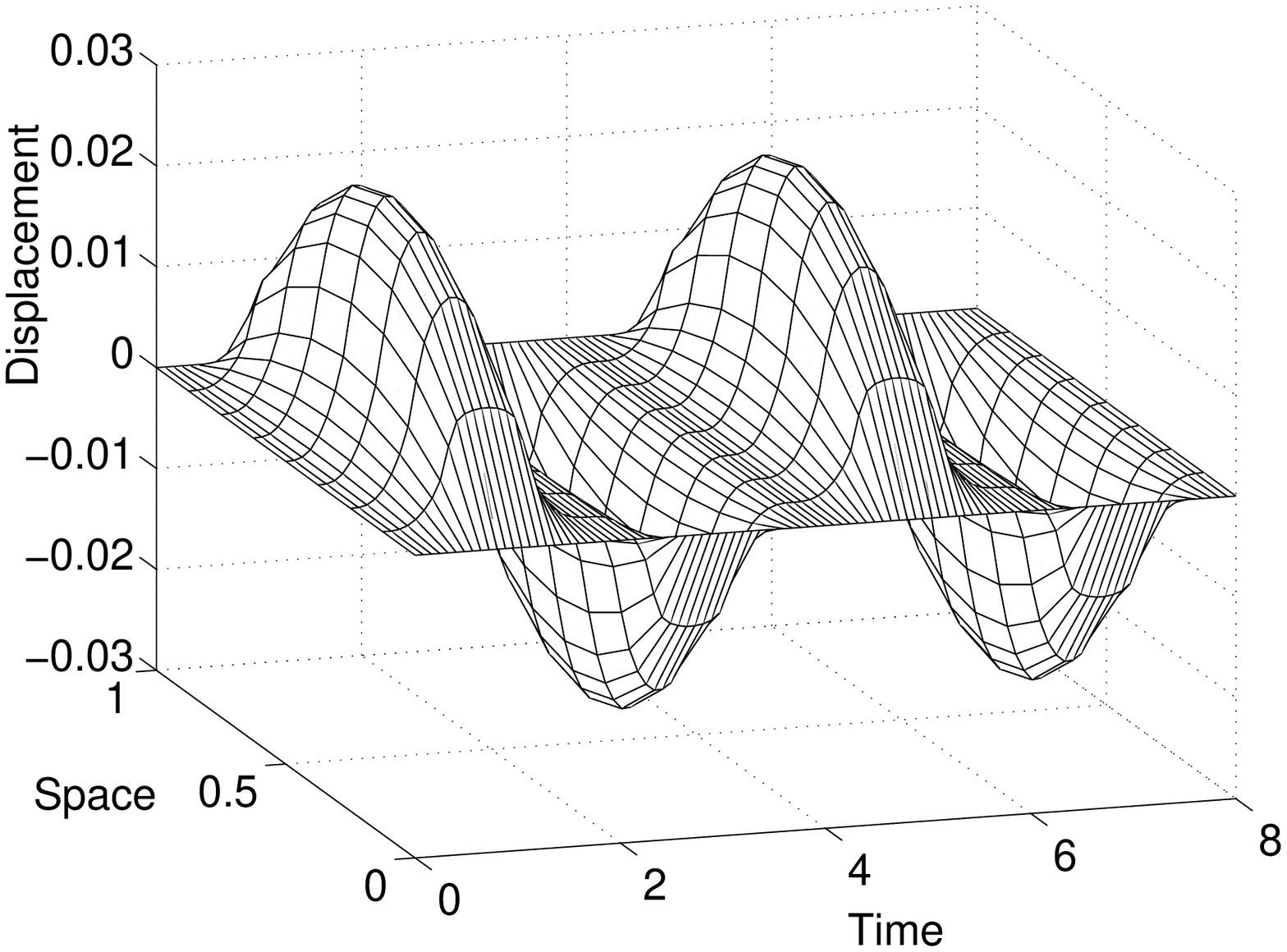}  \\
  \includegraphics[scale=0.4]{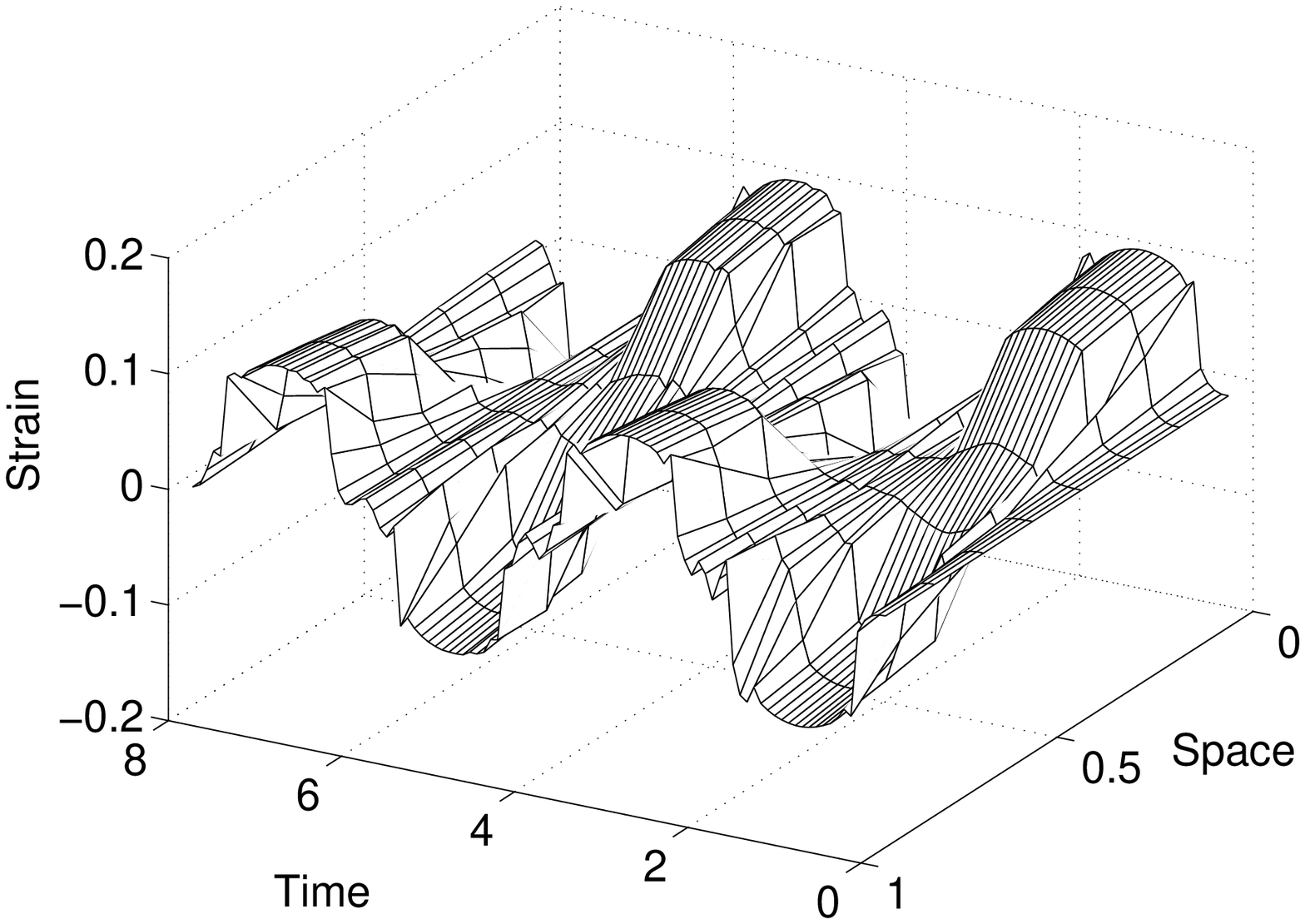}
  \includegraphics[scale=0.4]{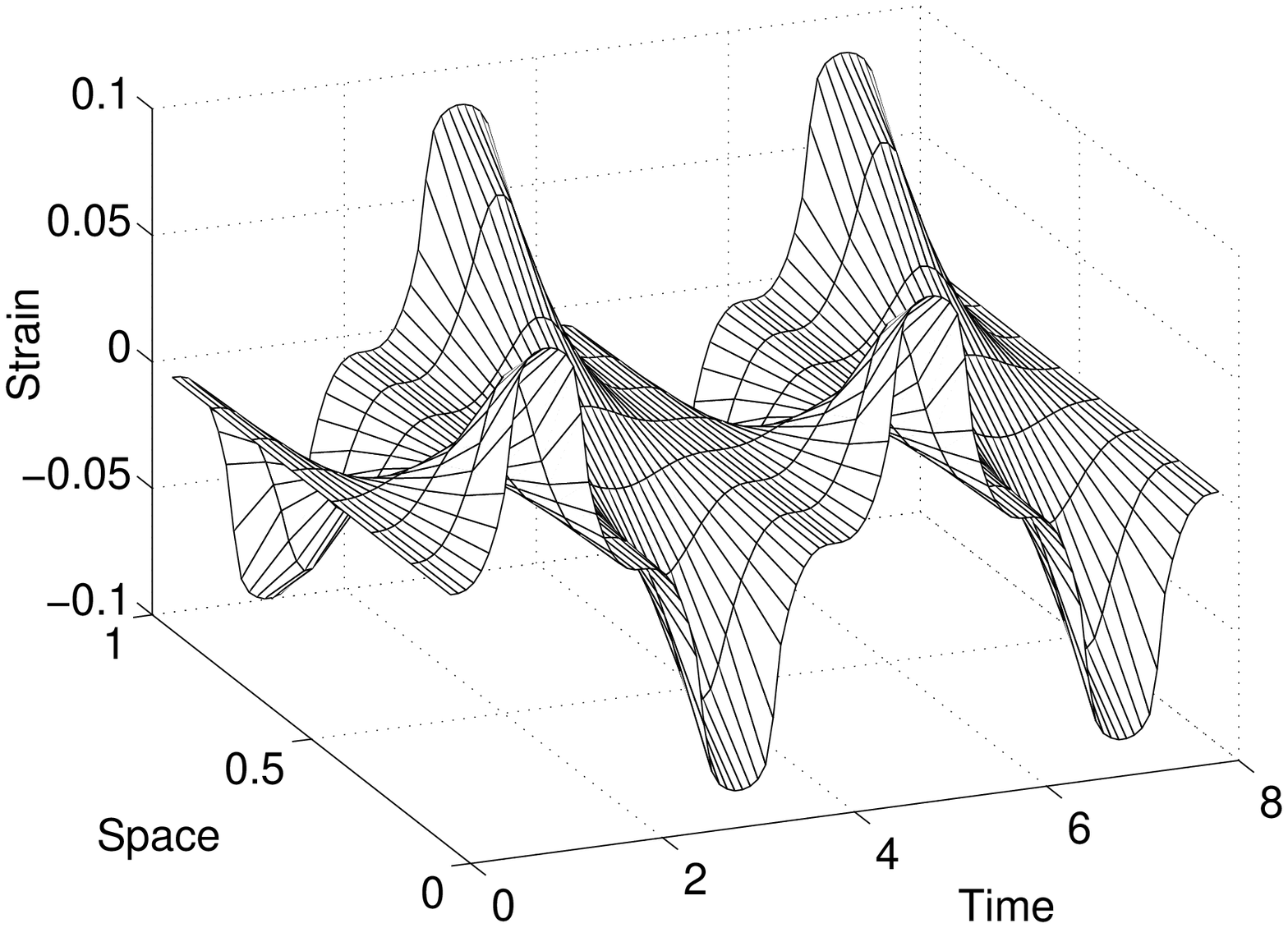}  \\
  \includegraphics[scale=0.4]{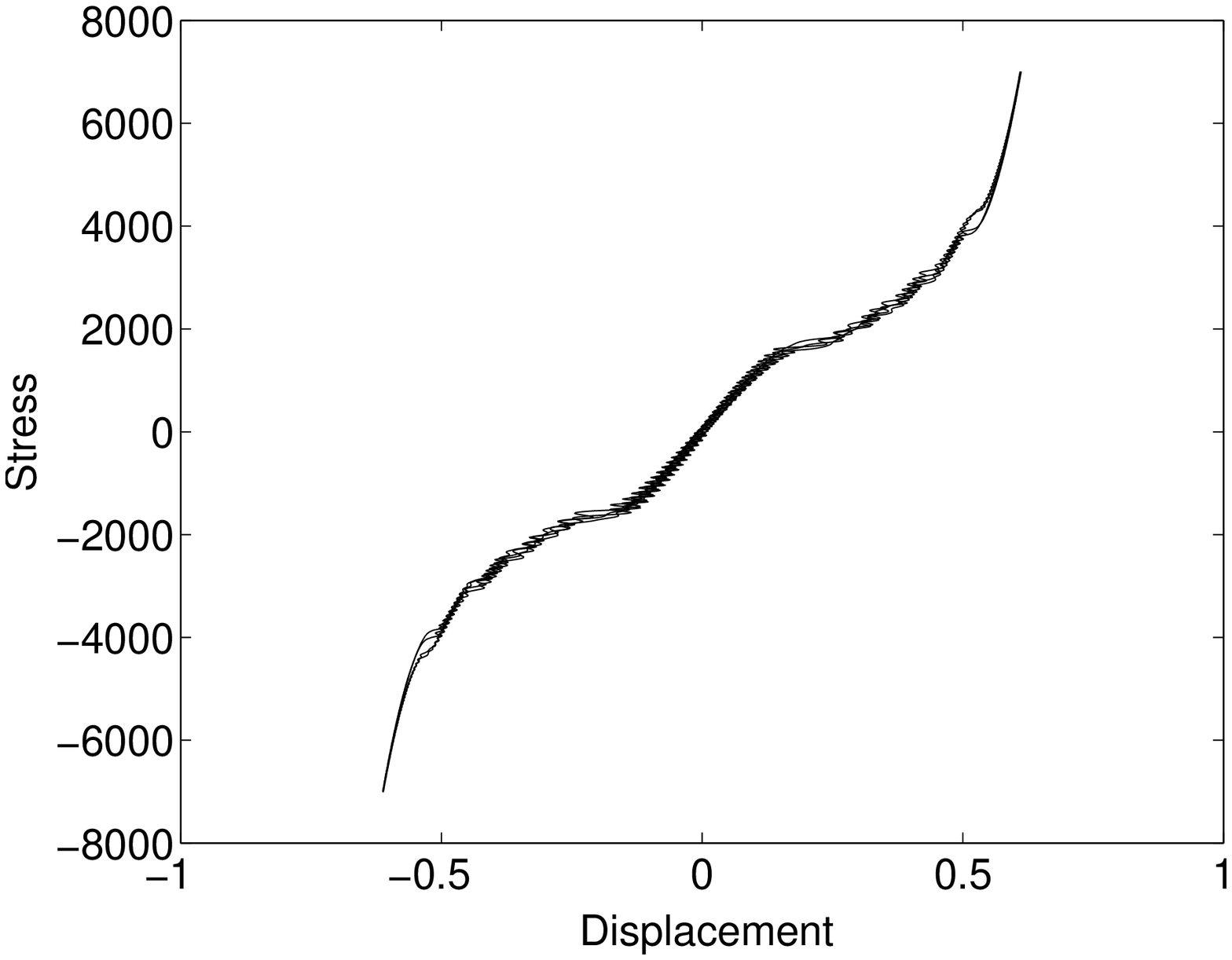}
  \includegraphics[scale=0.4]{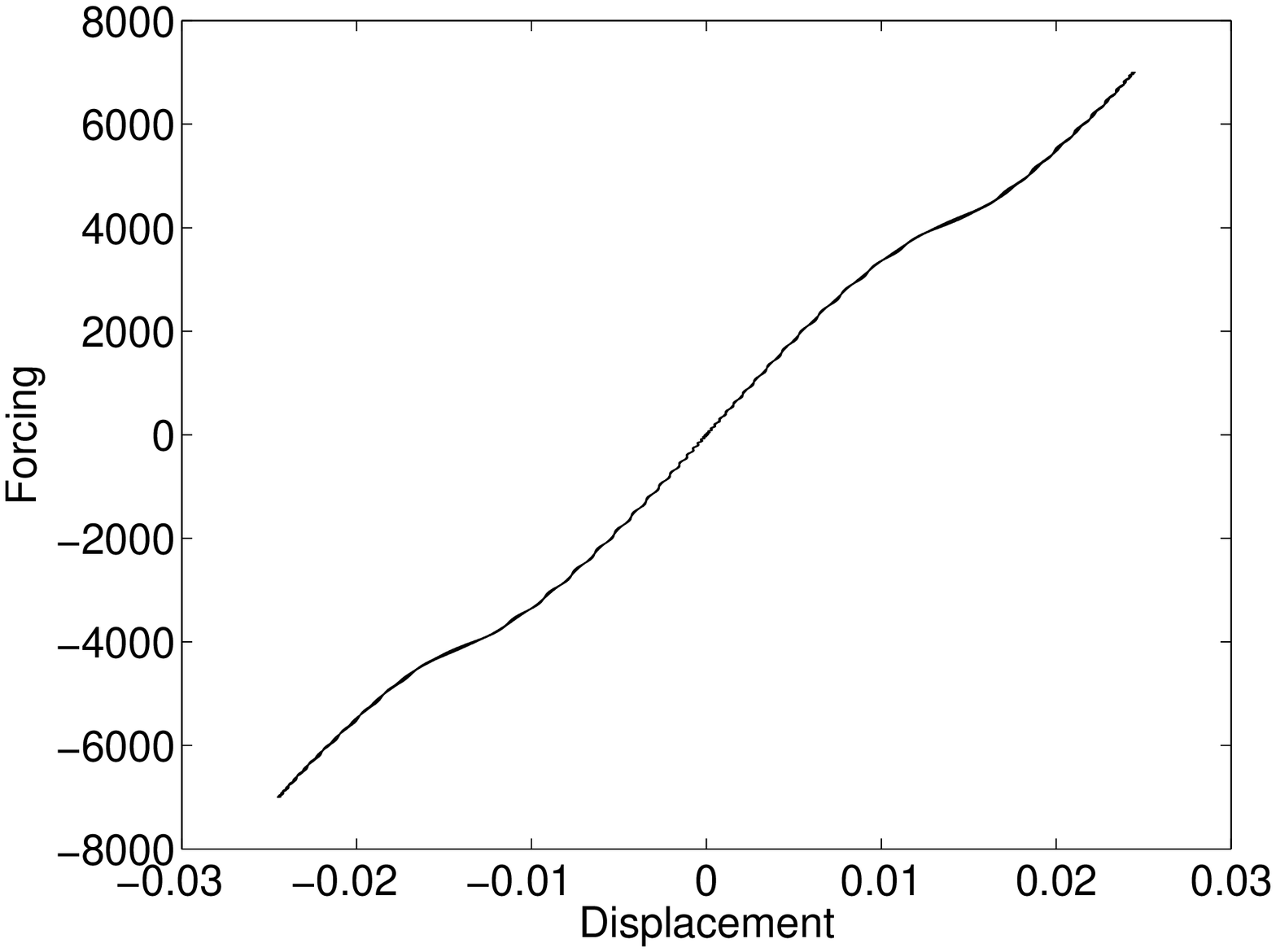}  \\
\caption{Dynamics and mechanical hysteresis of SMA at medium and high
temperature }
\end{figure}


\newpage

\begin{figure}
  \includegraphics[scale=0.4]{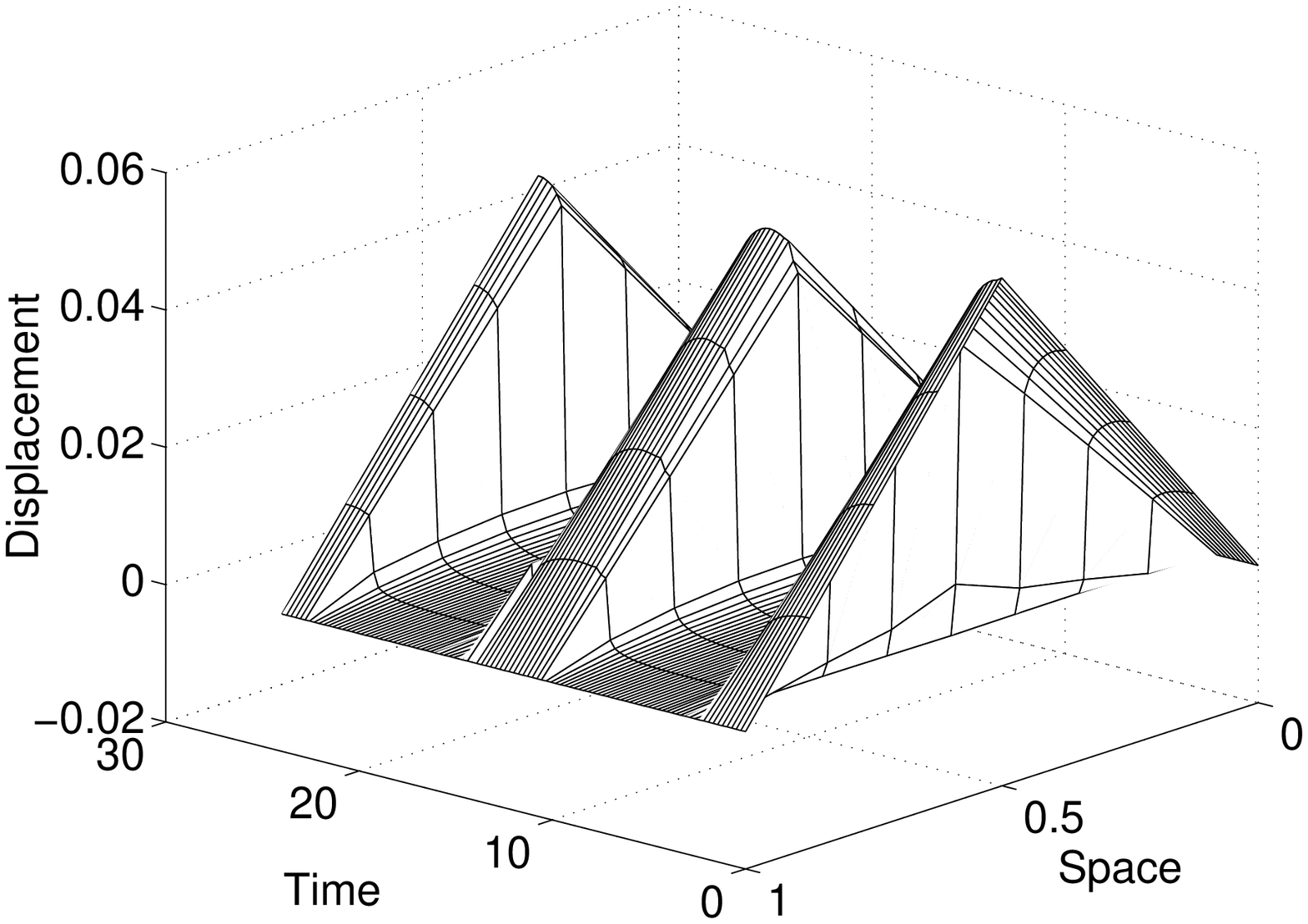}
  \includegraphics[scale=0.4]{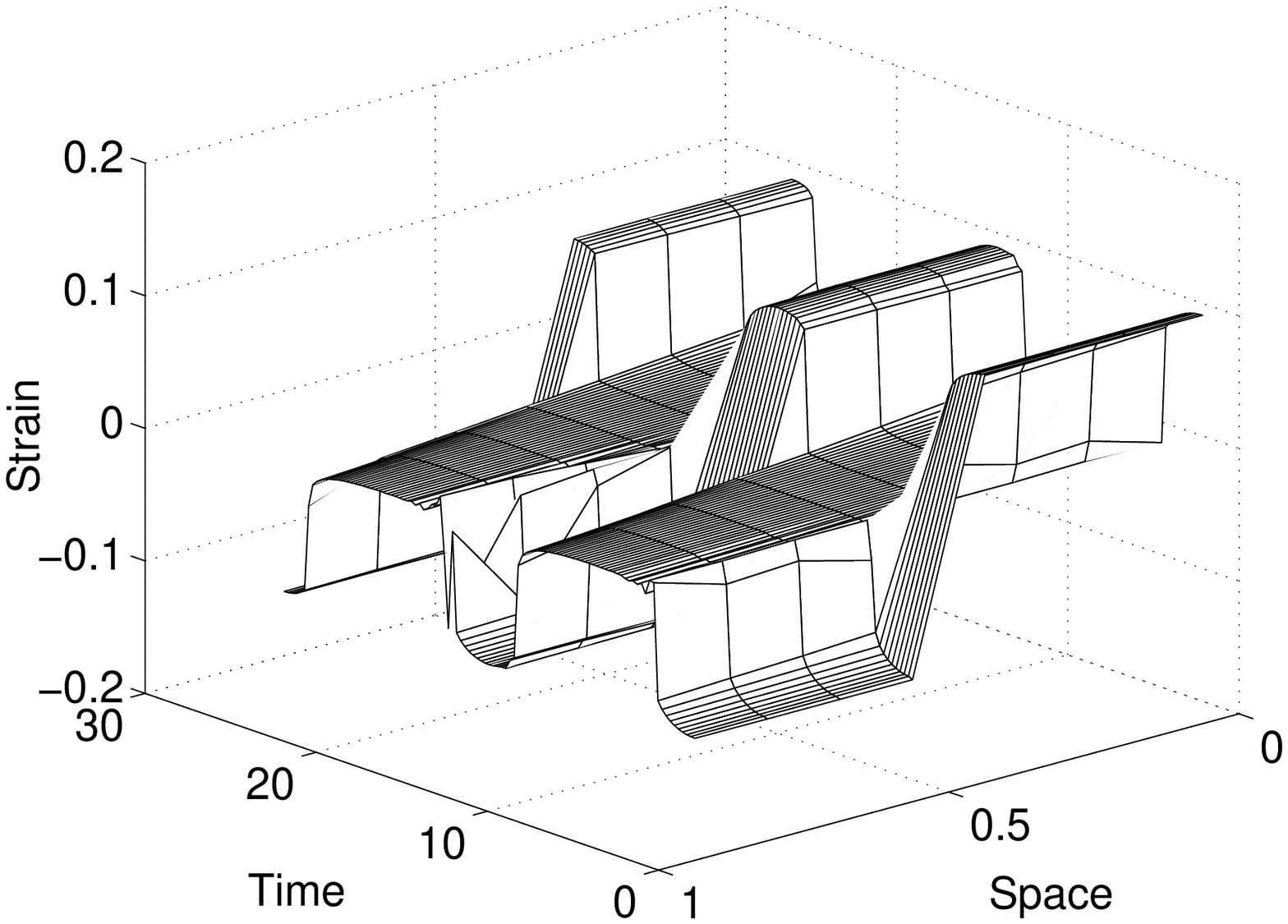}   \\
  \includegraphics[scale=0.4]{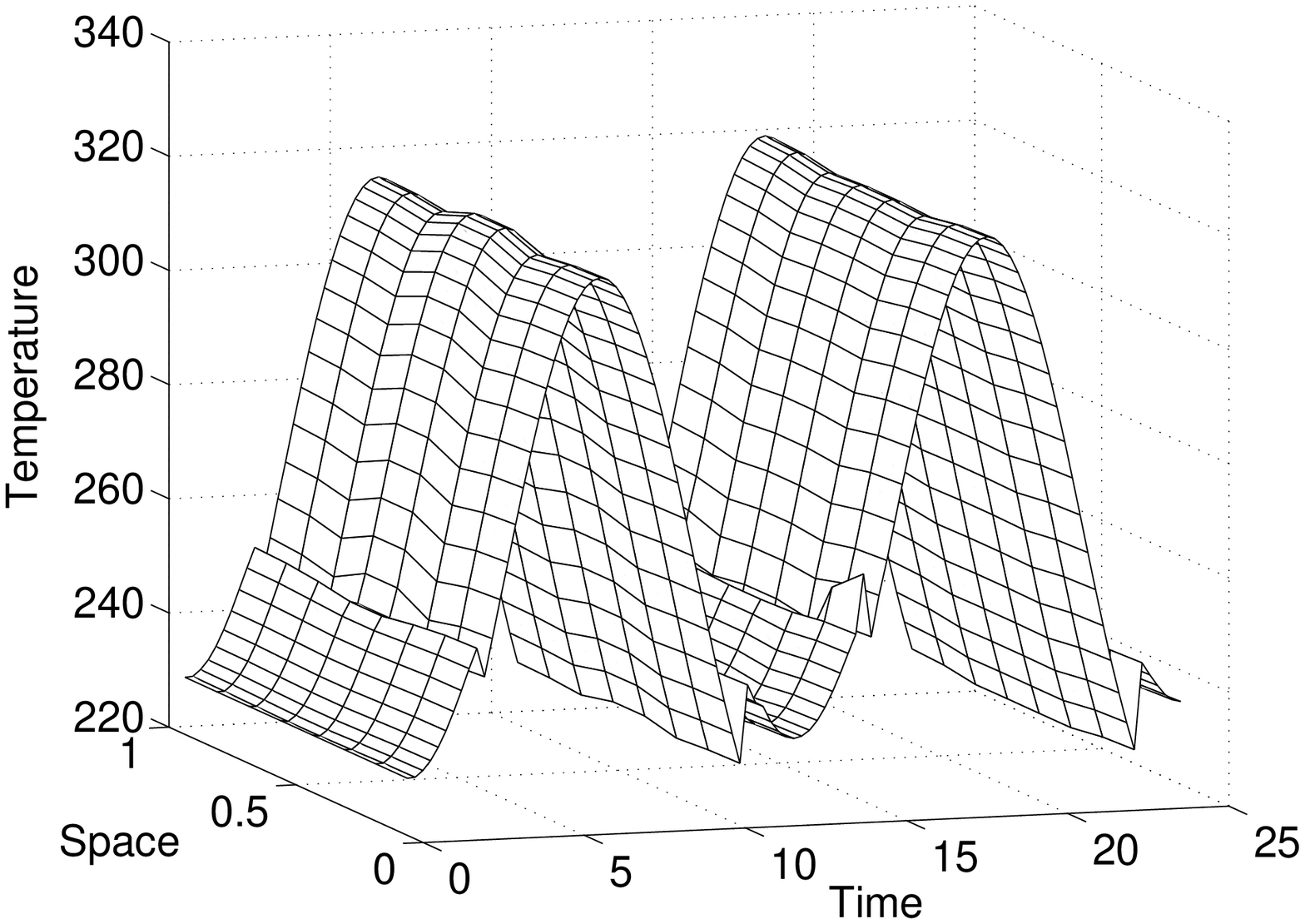}
  \includegraphics[scale=0.4]{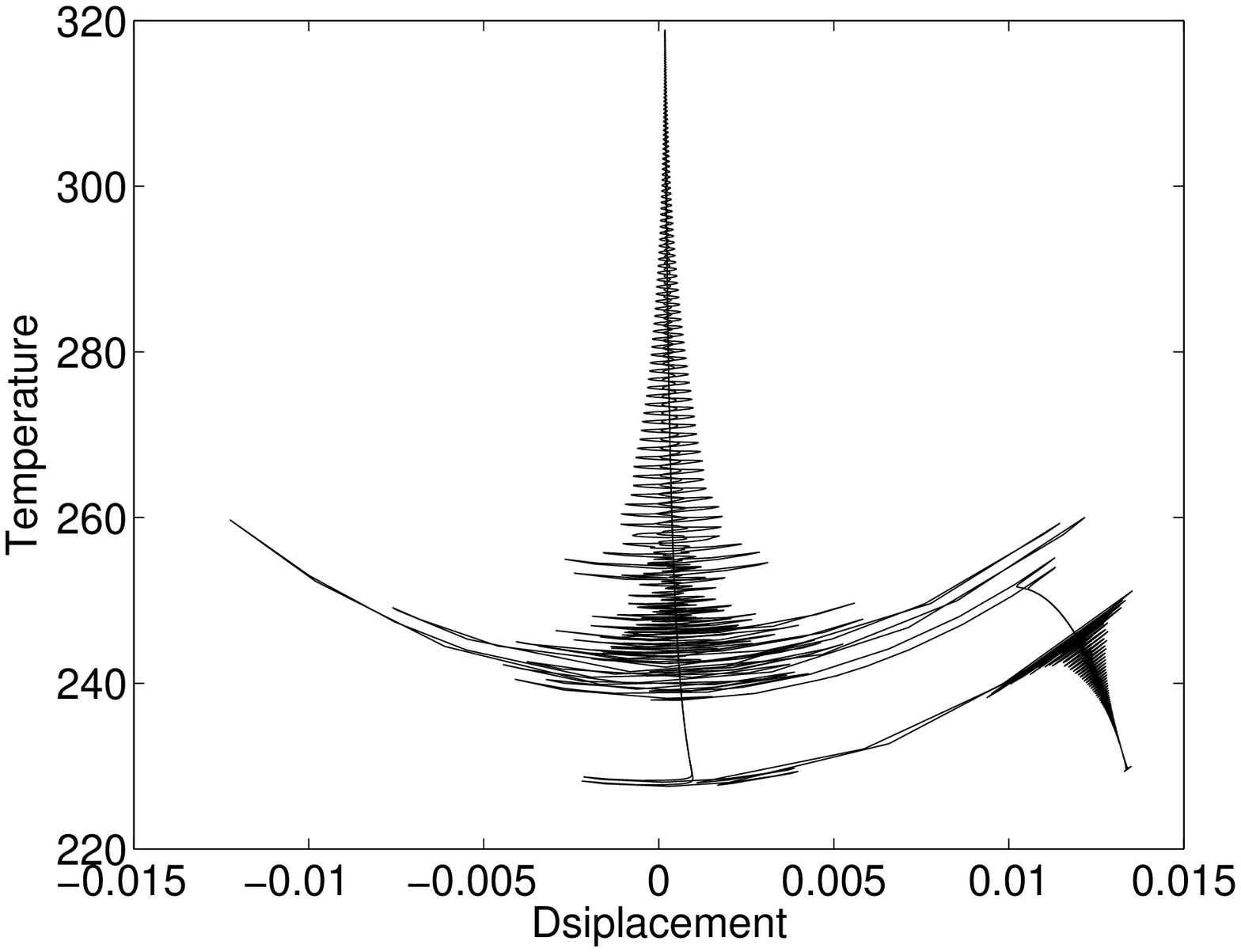}
\caption{Thermally induced phase transformation and thermal hysteresis in
SMA rod.}
\end{figure}


\newpage

\begin{figure}
{\epsfig{file=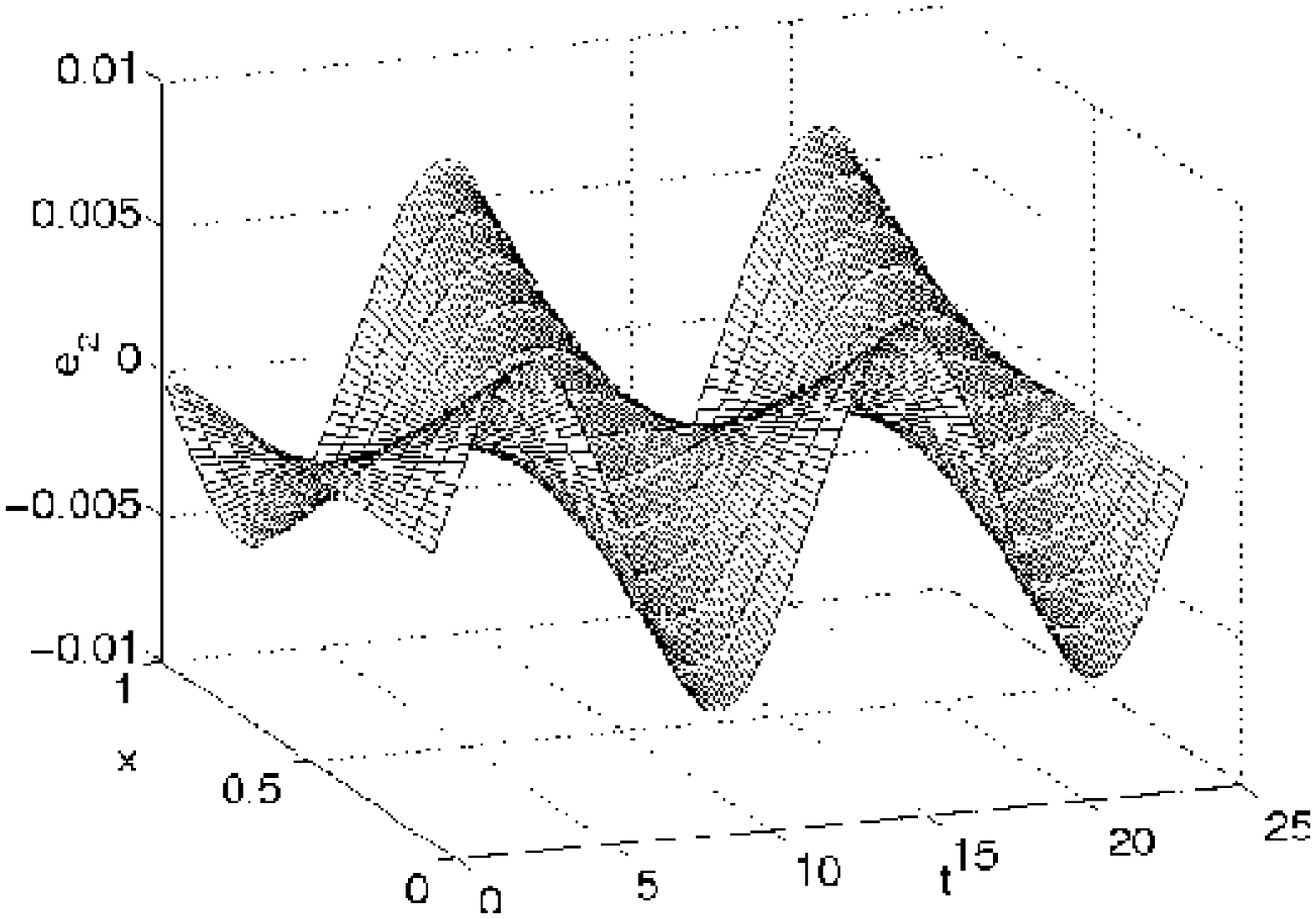, height = 5cm, width = 7cm } }
{\epsfig{file=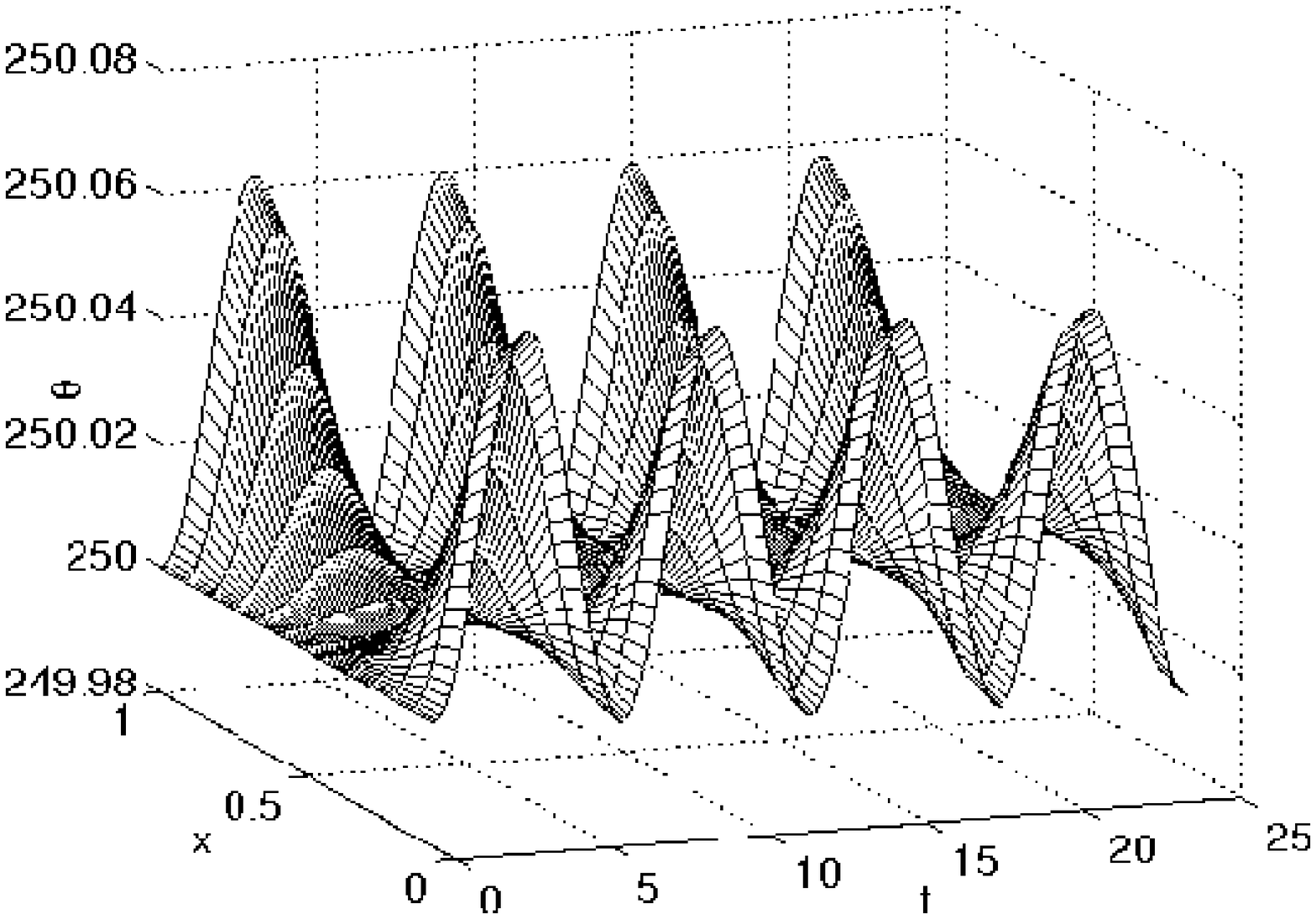, height = 5cm, width = 7cm } }    \\
{\epsfig{file=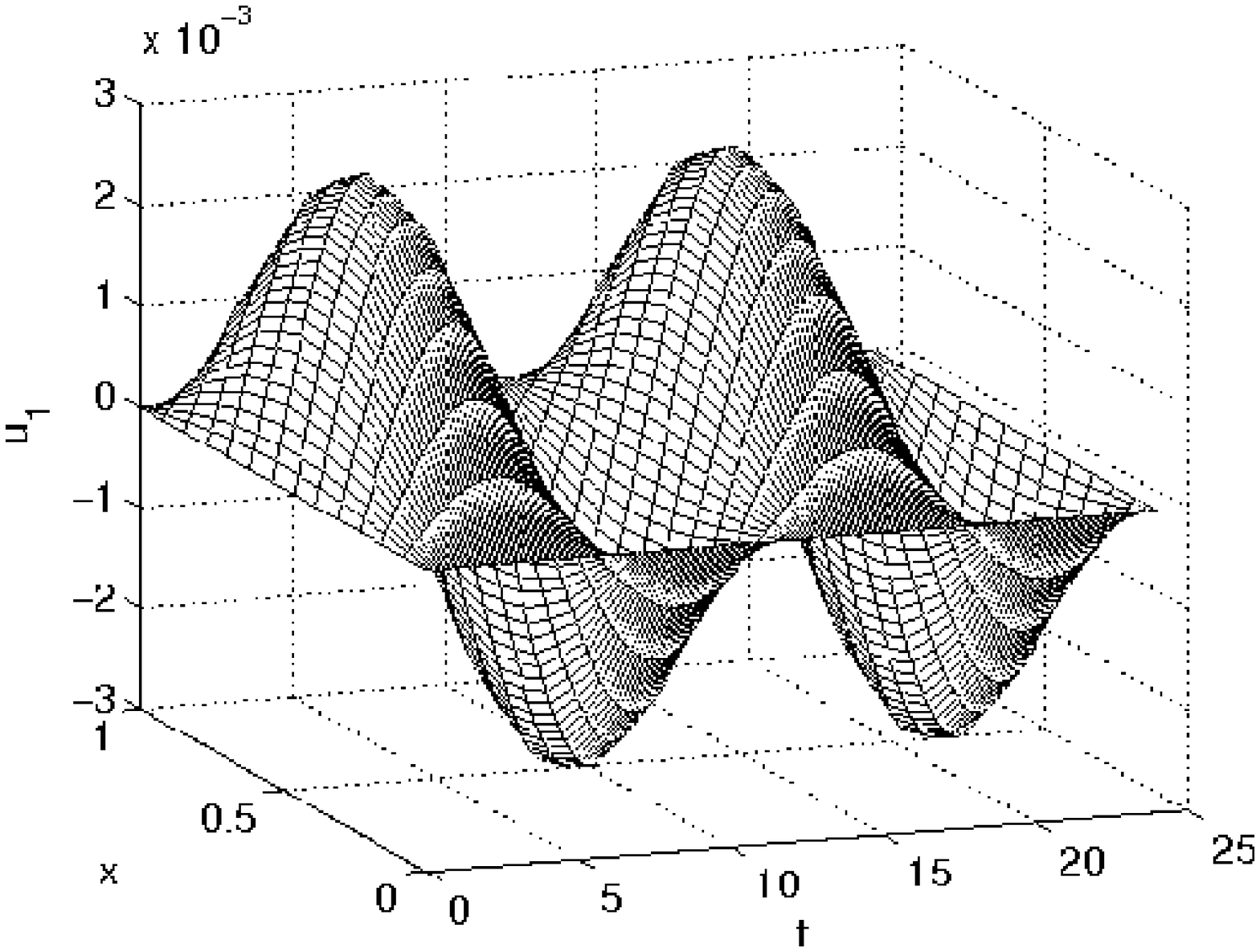, height = 5cm, width = 7cm } }
{\epsfig{file=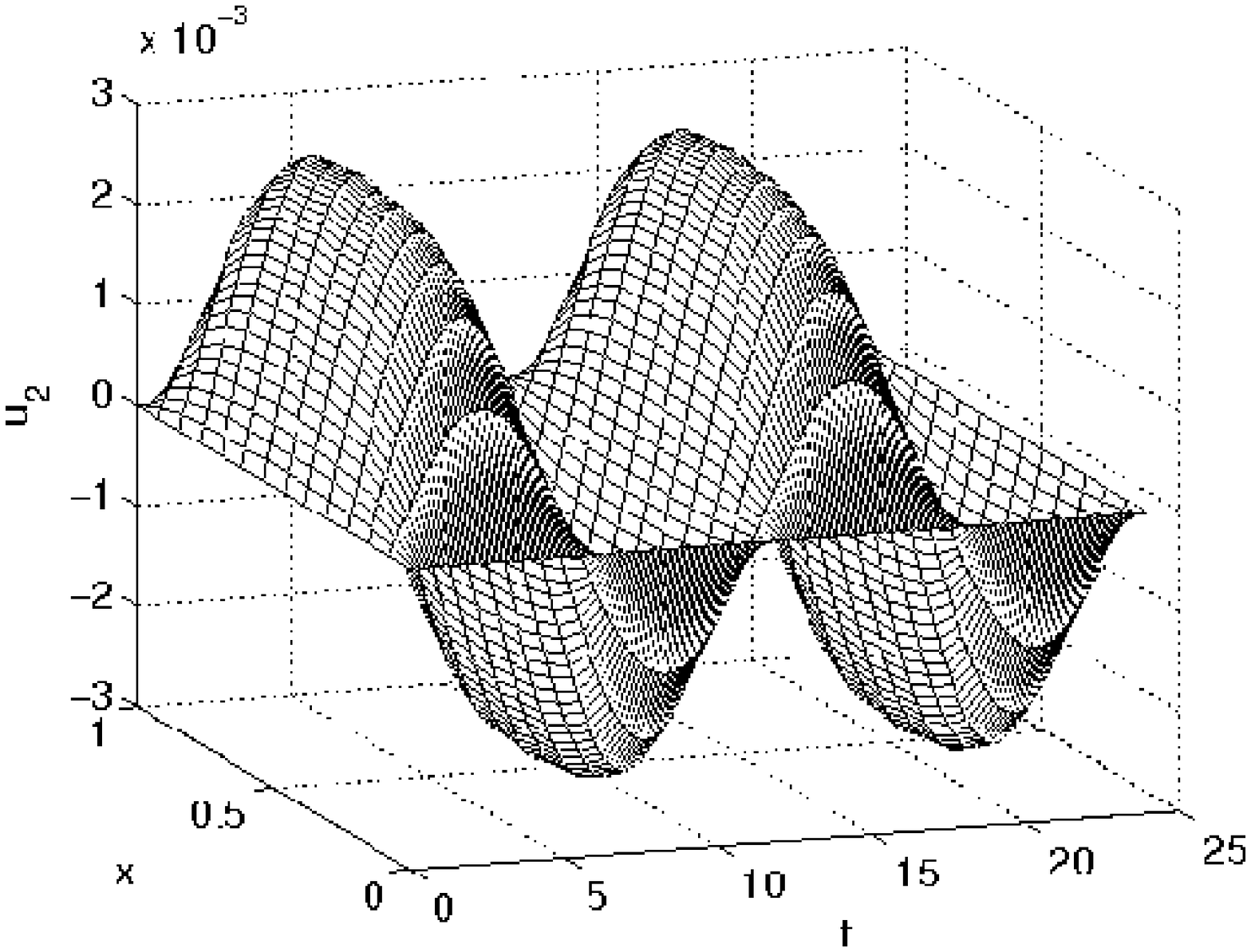, height = 5cm, width = 7cm } }    \\
{\epsfig{file=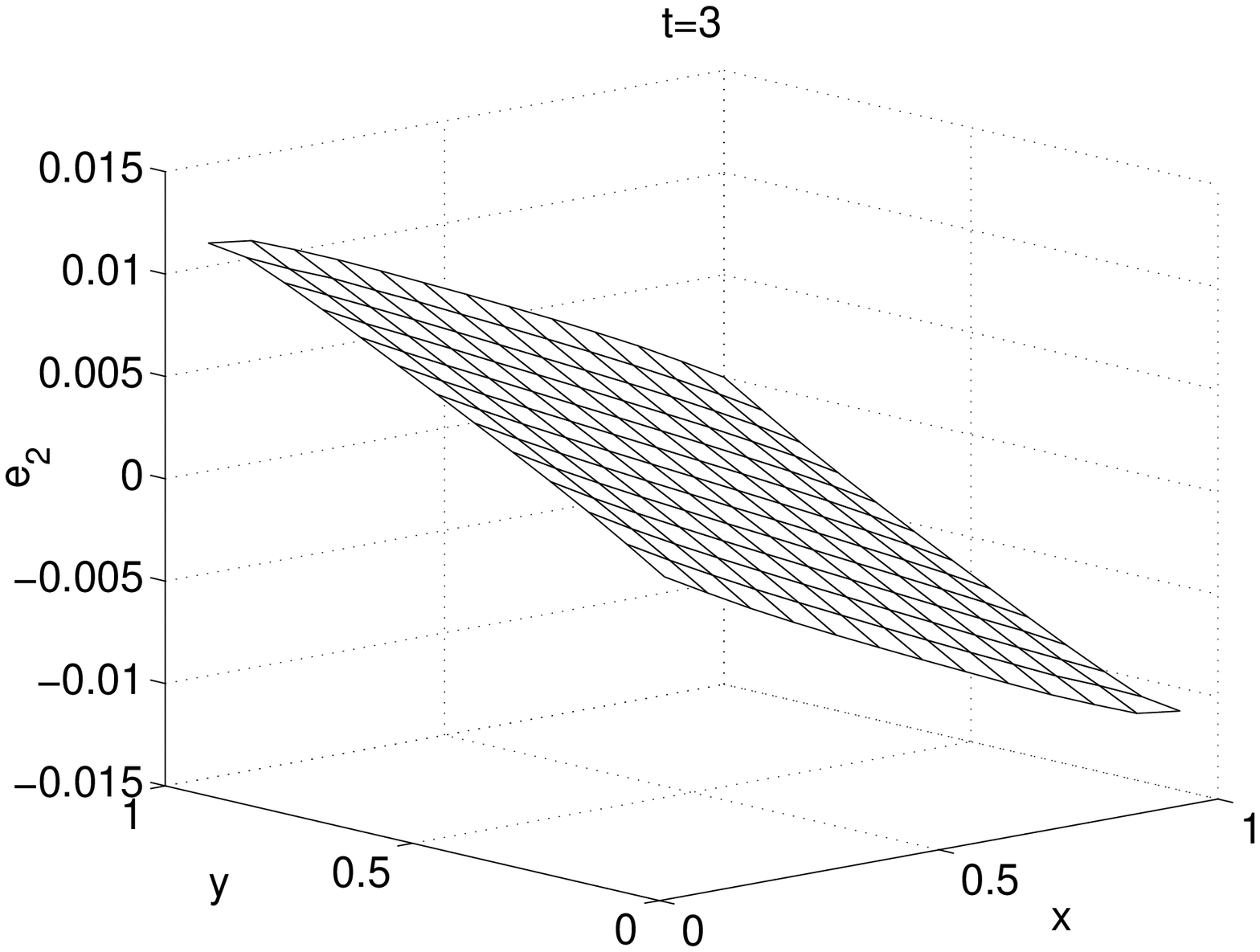, height = 5cm, width = 7cm } }
{\epsfig{file=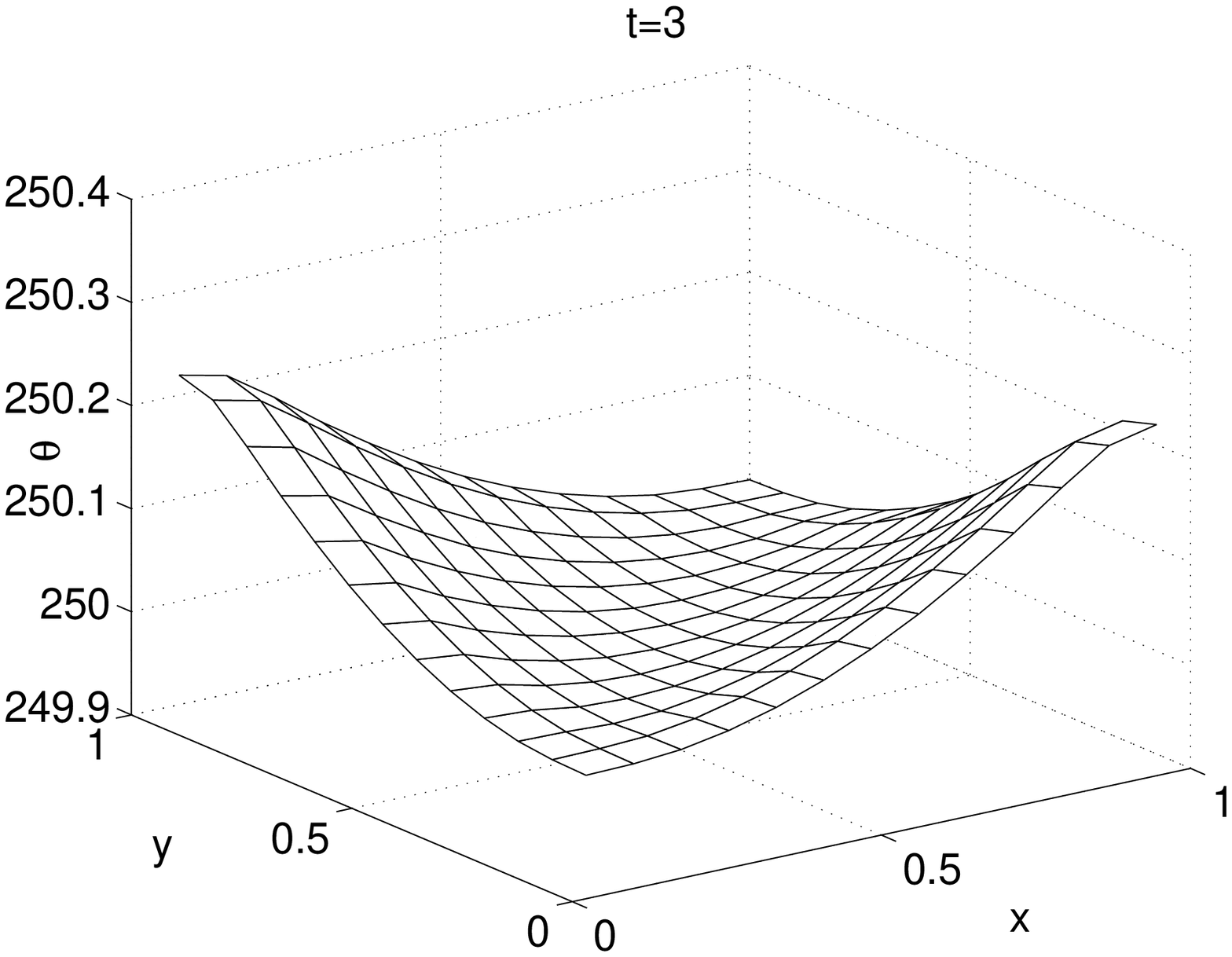, height = 5cm, width = 7cm } }    \\
{\epsfig{file=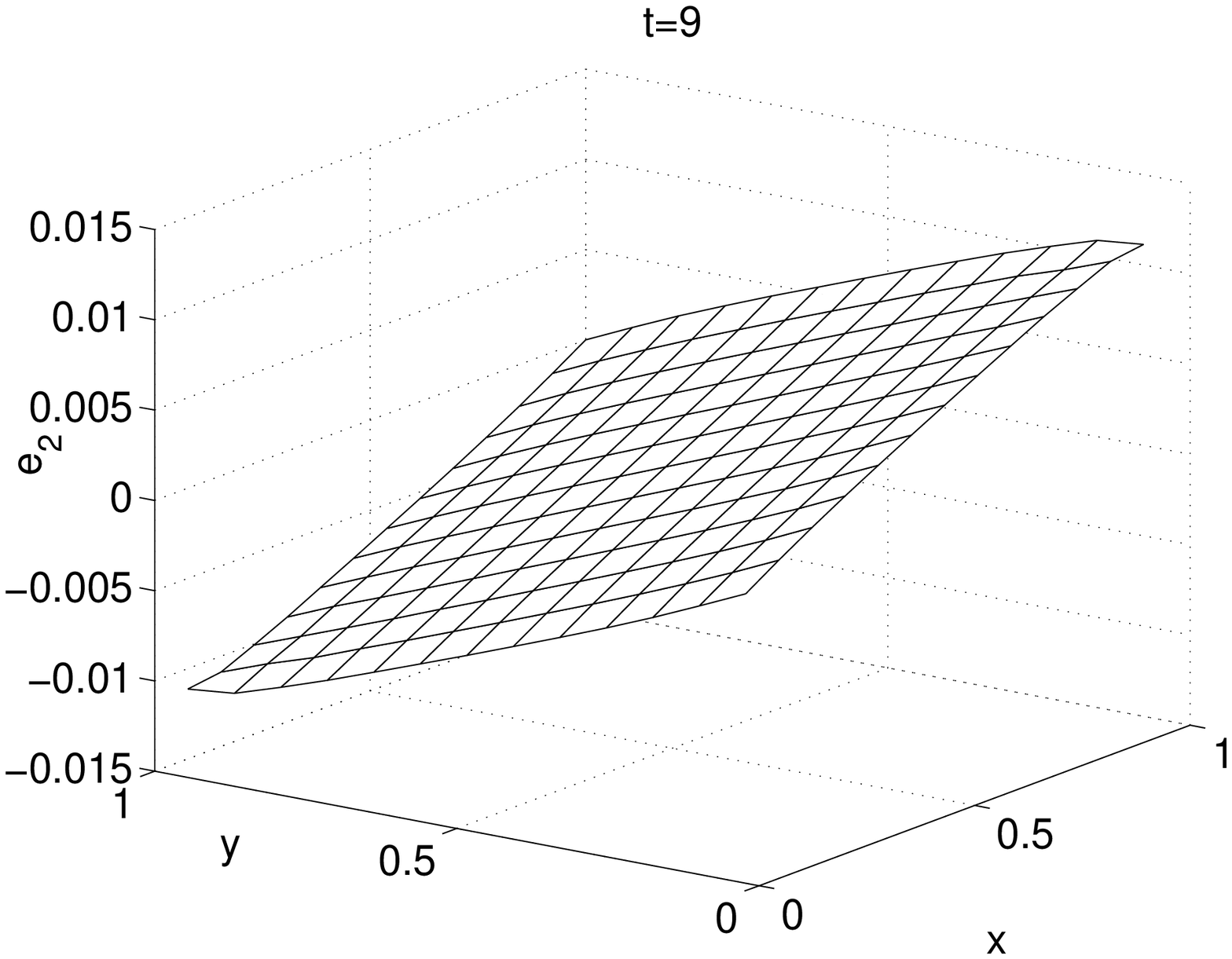, height = 5cm, width = 7cm } }
{\epsfig{file=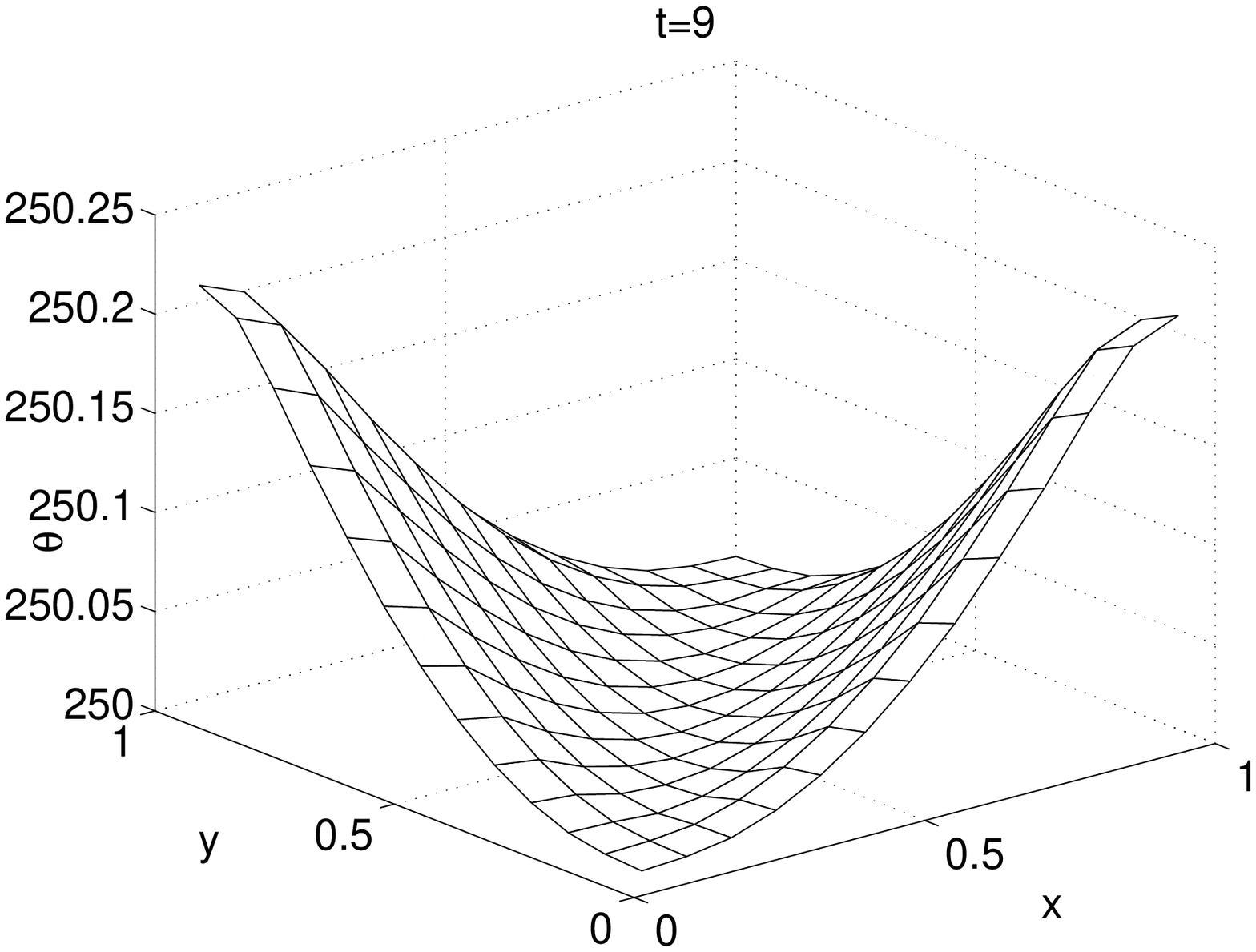, height = 5cm, width = 7cm } }    \\
\caption{Thermo-mechanical waves in SMA patches caused by varying mechanical
loadings.}
\end{figure}


\newpage

 \begin{figure}

{\epsfig{file=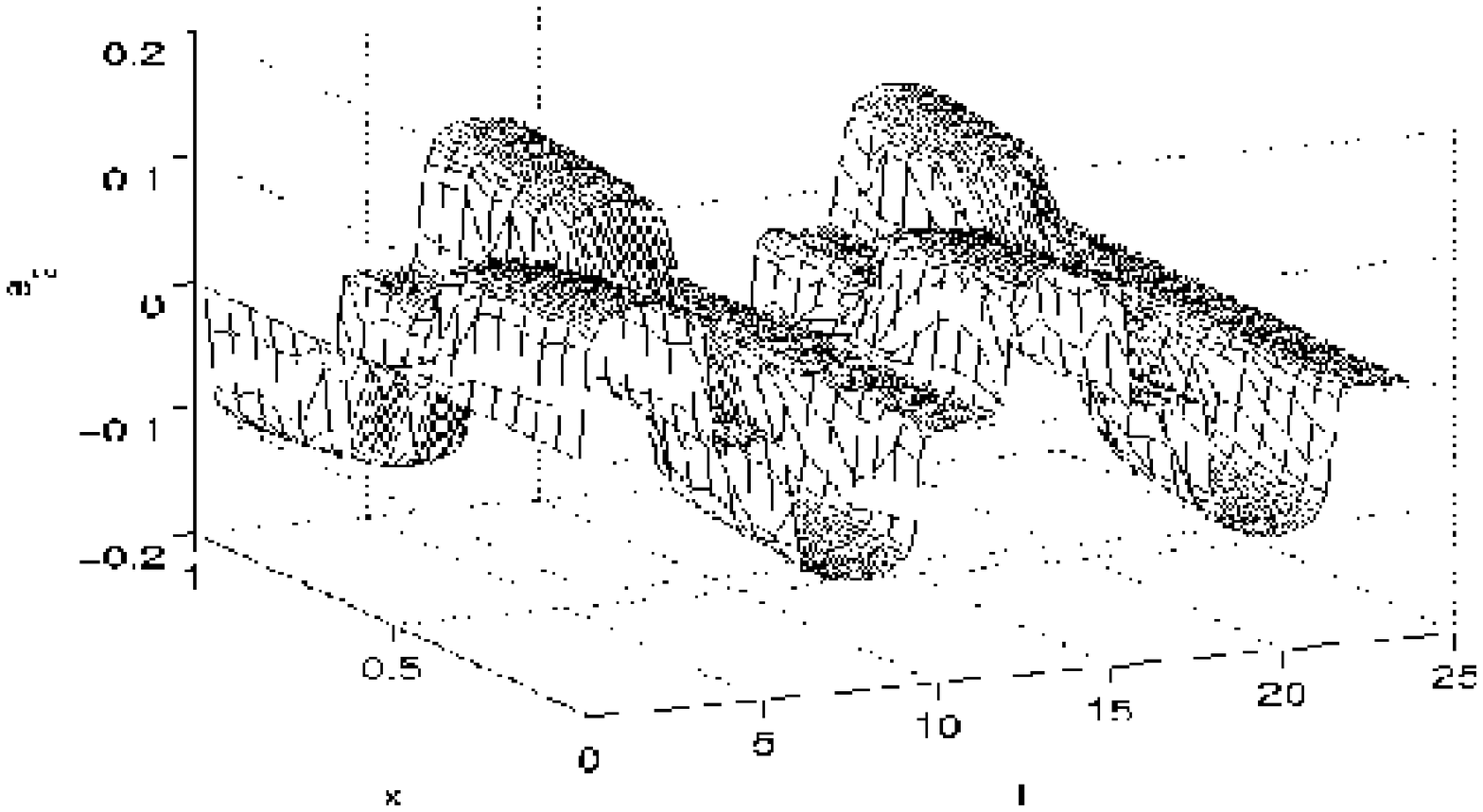, height = 8cm, width = 7cm } }
{\epsfig{file=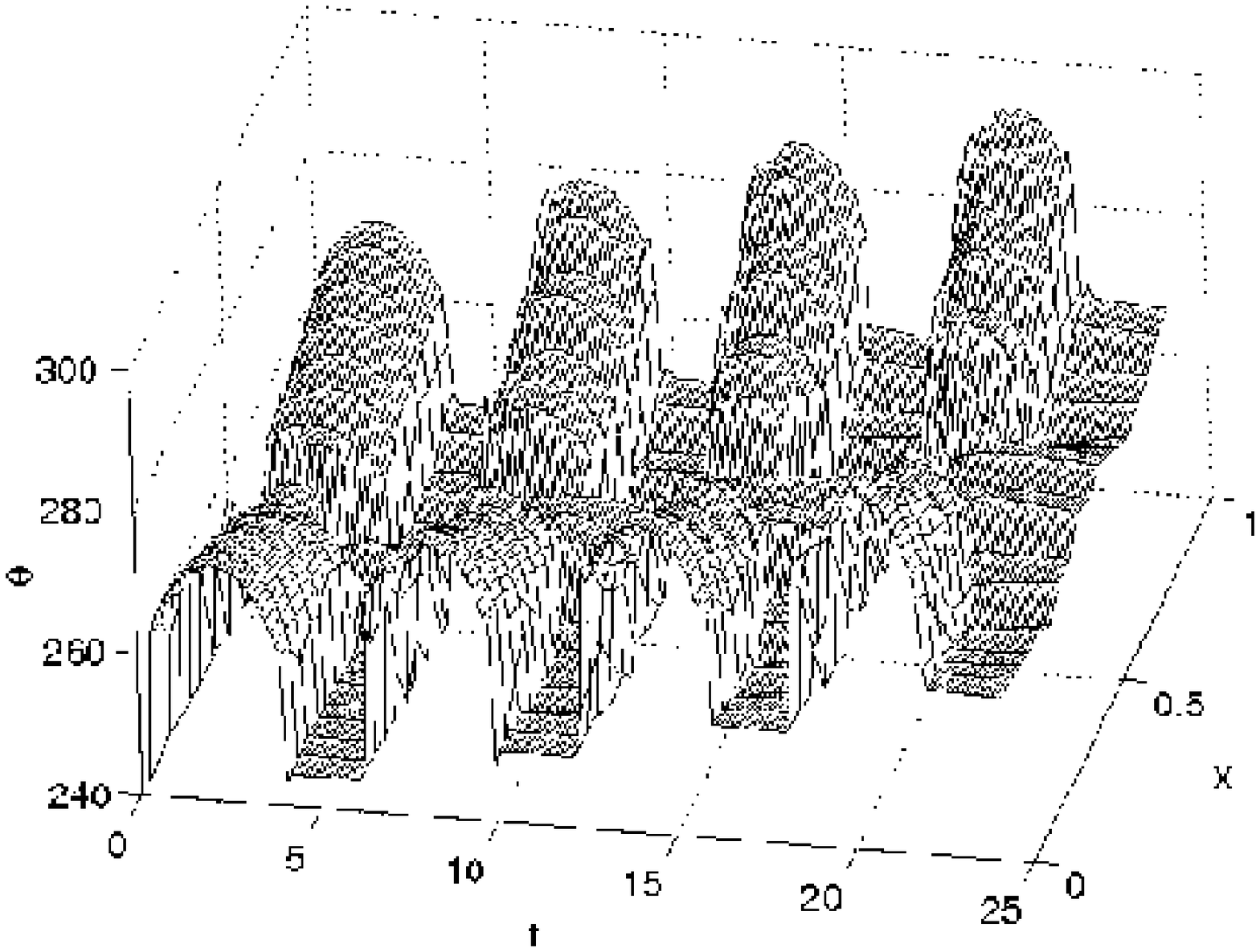, height = 8cm, width = 7cm } }   \\
{\epsfig{file=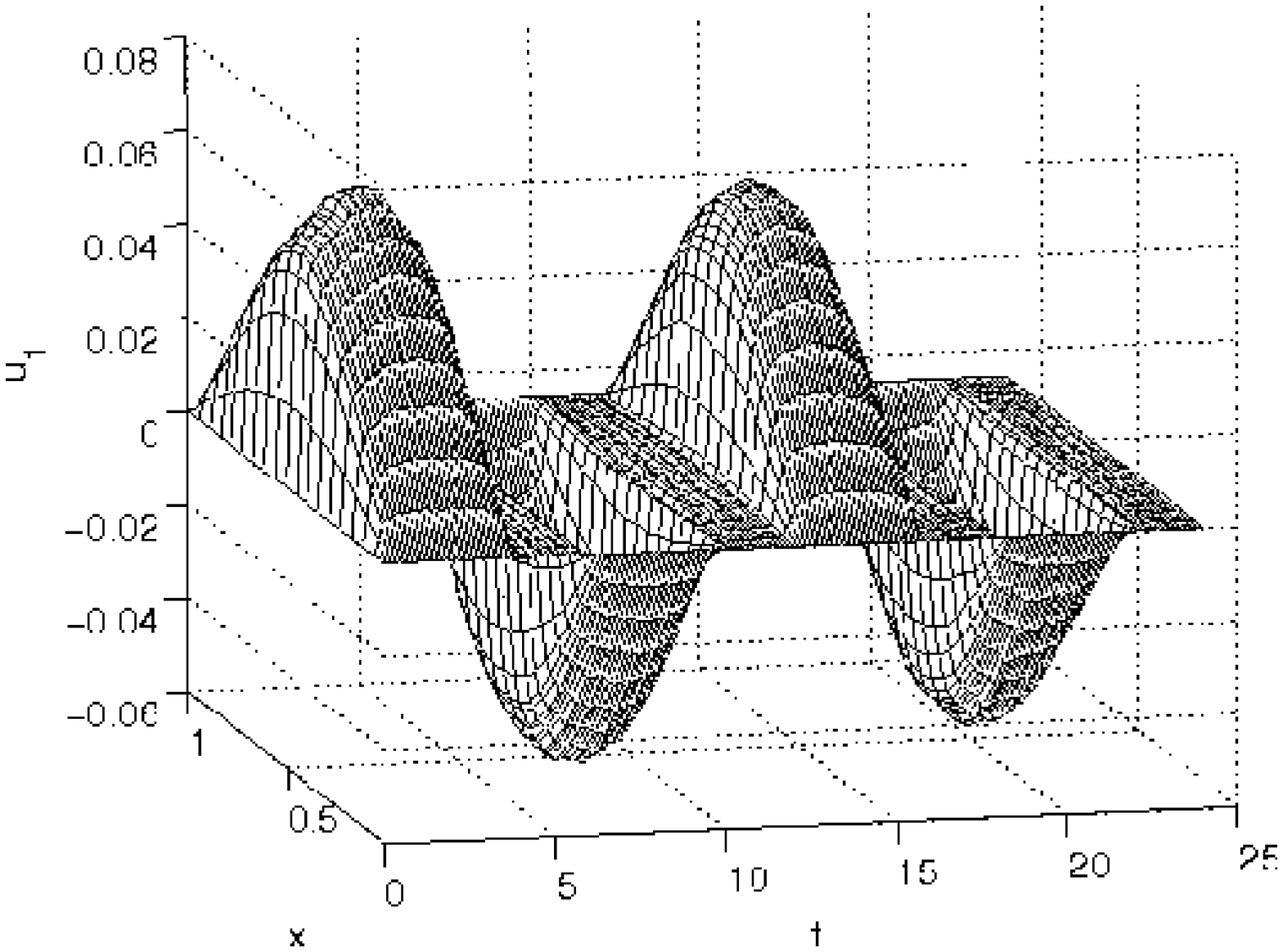, height = 8cm, width = 7cm } }
{\epsfig{file=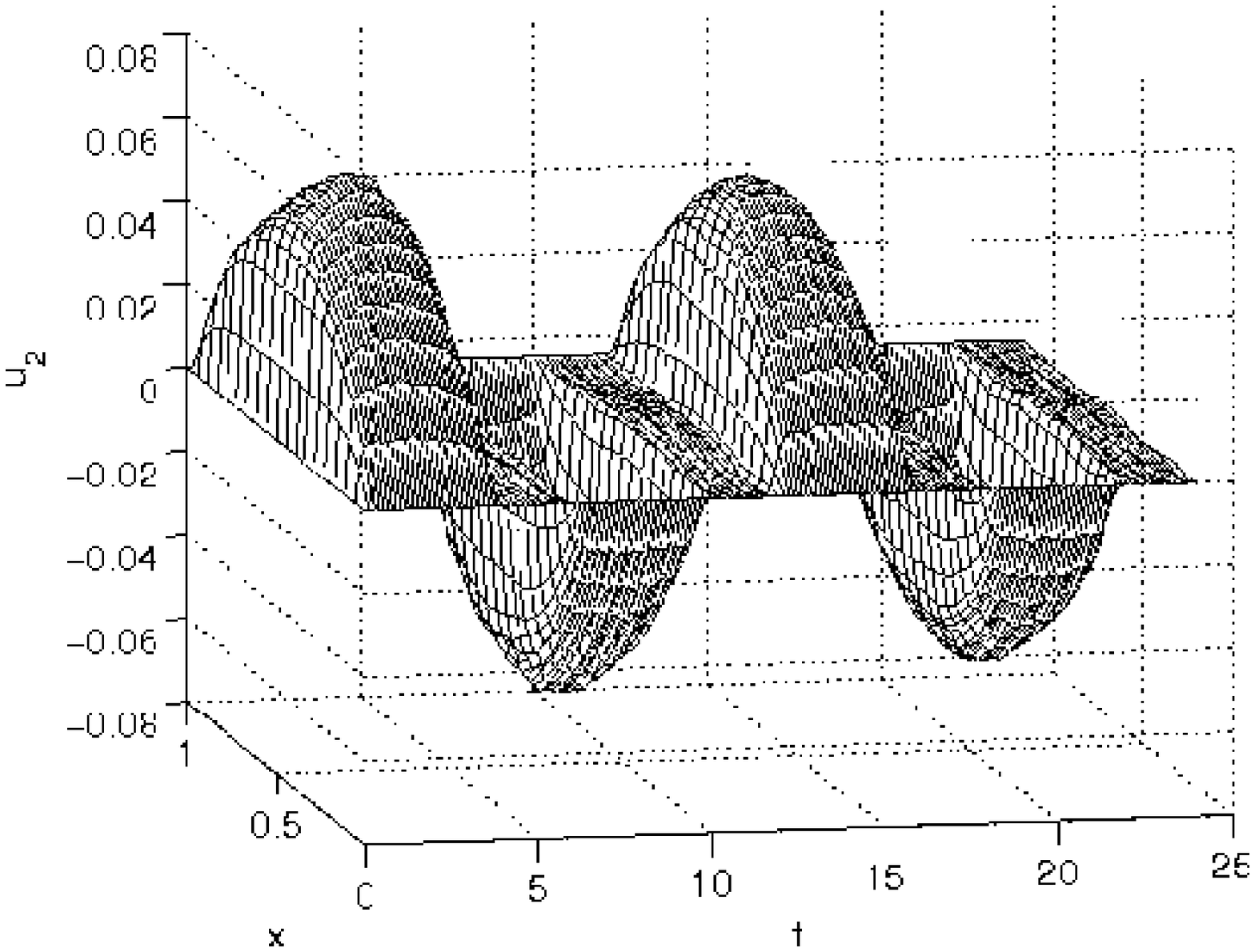, height = 8cm, width = 7cm } }   \\
 \caption{Martensitic transformation in a square SMA patch caused by
mechanical loadings in the $x$ and $y$ direction. }
 \end{figure}


\newpage

\begin{figure}
{\epsfig{file=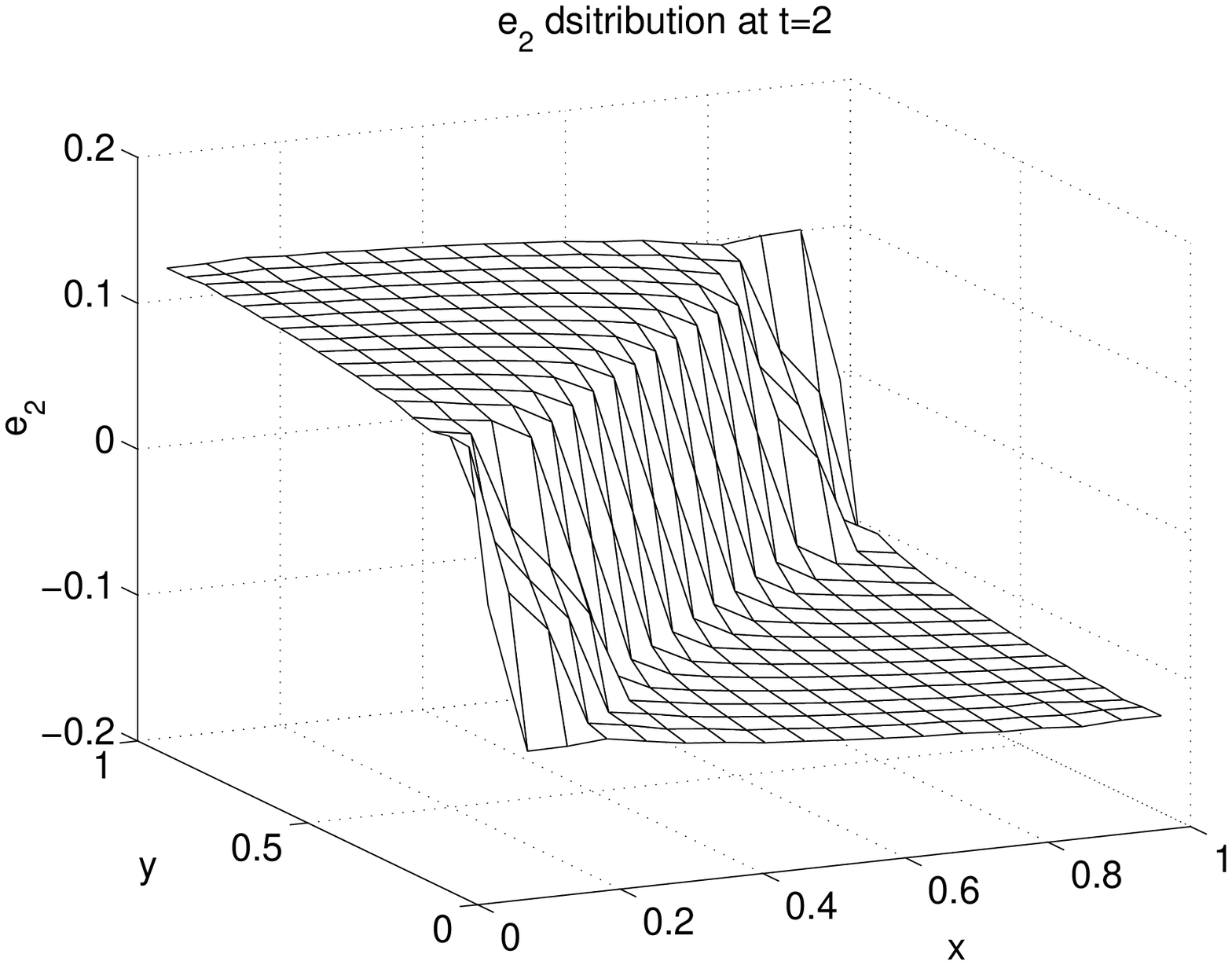, height = 8cm, width = 7cm } }
{\epsfig{file=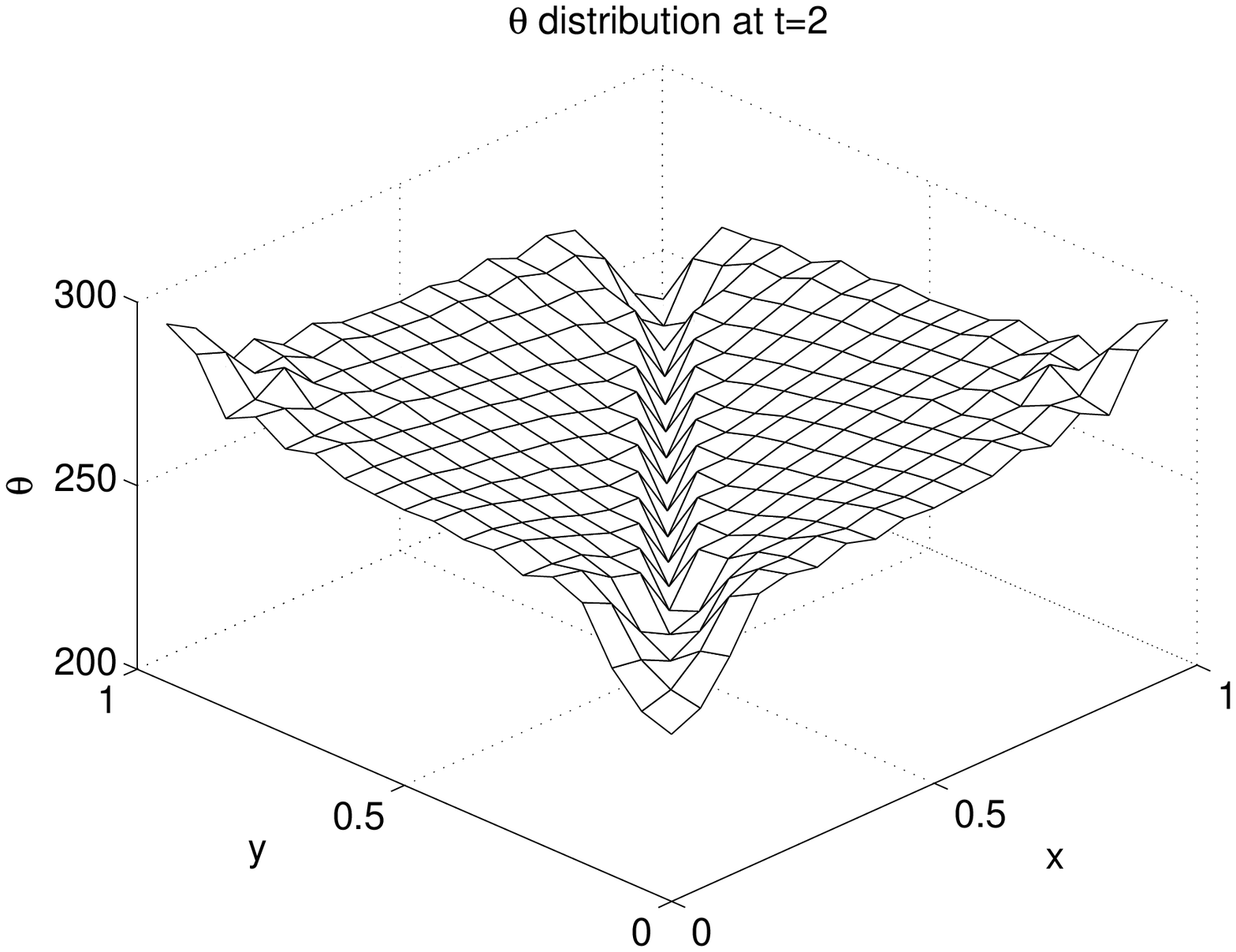, height = 8cm, width = 7cm } }   \\
{\epsfig{file=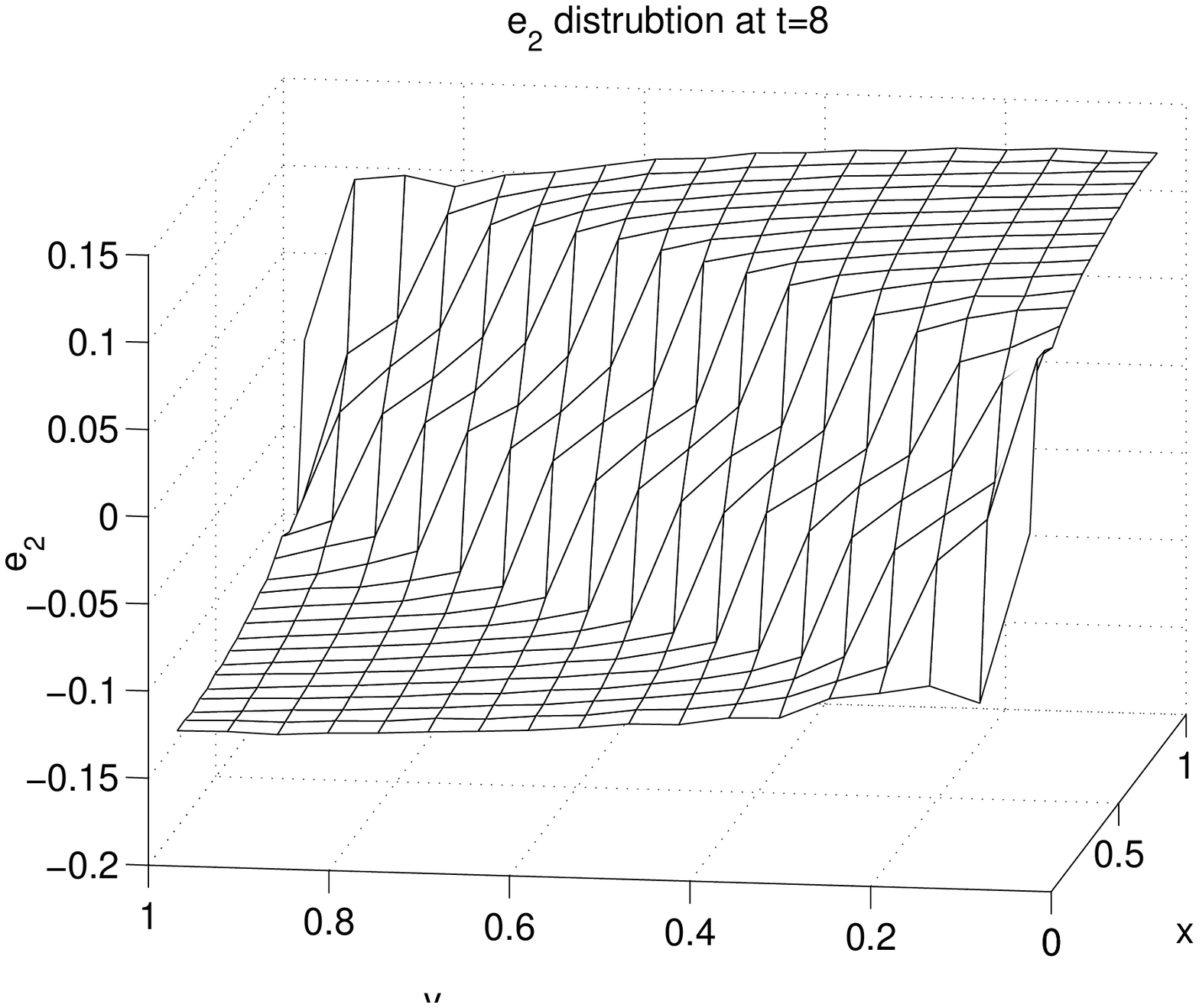, height = 8cm, width = 7cm } }
{\epsfig{file=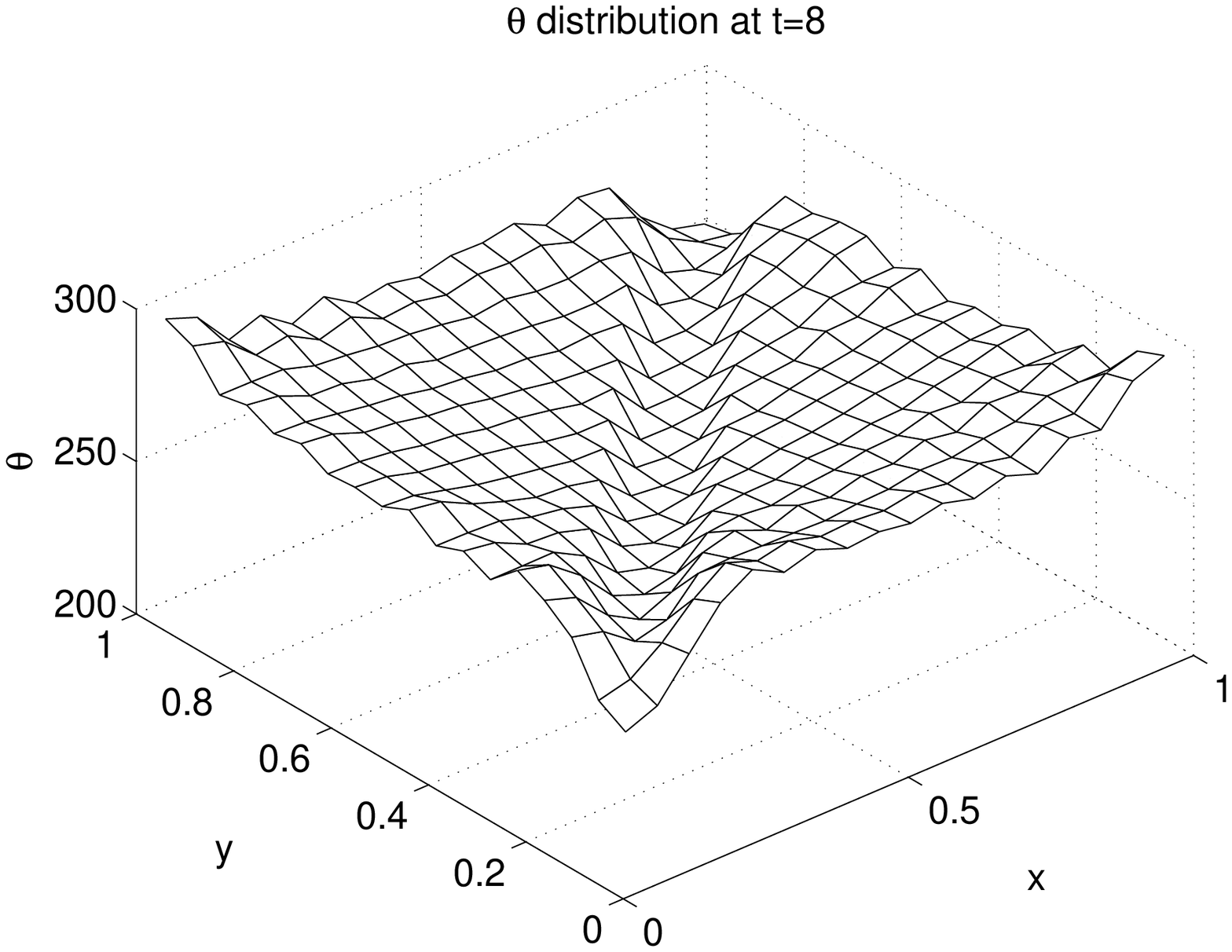, height = 8cm, width = 7cm } }   \\
\caption{Martensite combinations and temperature distributions in a square SMA
patch after phase transformations caused by mechanical loadings in the $x$ and
$y$ direction. }
\end{figure}


\end{document}